\documentclass[12pt]{article}
\begin{document}
\begin {titlepage}
\begin{flushright} ULB--TH--01/21\\   Bicocca FT-01-14 \\ DCPT-01/55\\
hep-th/0106235\\
\end{flushright}
\begin{center} {\large \bf  Brane fusion in the bosonic string and the
emergence of fermionic strings }\\
\vspace{.3cm}  Fran\c cois Englert,${}^{a,}$\footnote{ E-mail :
fenglert@ulb.ac.be} Laurent Houart${}^{b,}$\footnote{ E-mail :
Laurent.Houart@mib.infn.it} and Anne Taormina${}^{c,}$\footnote{ E-mail :
anne.taormina@durham.ac.uk}\\
\vspace{.3cm} ${}^a${\it Service de Physique Th\'eorique}\\ {\it Universit\'e
Libre de Bruxelles, Campus Plaine, C.P.225}\\ {\it Boulevard du Triomphe,
B-1050 Bruxelles, Belgium}\\
\vspace{.2cm}
 ${}^b${\it Dipartimento di Fisica }\\  {\it Universit{\`a} Milano Bicocca}\\
{\it Piazza della Scienza 3, Milano, Italy.}\\
\vspace{.2cm}
 ${}^c${\it Department of Mathematical Sciences }\\  {\it University of
Durham}\\ {\it South Road, DH1 3LE Durham, England}\\
\end{center}
\begin{abstract}

 We review the emergence of the ten-dimensional fermionic
closed string theories from subspaces of the Hilbert space  of the
26-dimen-sional  bosonic closed string theory compactified on an
$E_8\times SO(16)$ lattice. They arise from a consistent truncation
procedure which generates    space-time fermions out of bosons. This
procedure  is   extended to  open string sectors. We prove that
truncation  of the unique tadpole-free $SO(2^{13})$   bosonic string
theory compactified on the above lattice determines  the
anomaly free Chan-Paton group of the  Type I theory and the
consistent Chan-Paton groups of Type O  theories. It
also predicts the tension of space-filling  D-branes in
these fermionic theories. The derivation of these   fermionic
string properties  from  bosonic  considerations alone points towards a
dynamical origin of the  truncation  process.
Space-time fermions and supersymmetries would then arise from
bosonic degrees of freedom and no fermionic degrees of freedom would
be needed in a fundamental theory of quantum gravity.

\end{abstract}
\end {titlepage}
\addtocounter{footnote}{-3}
\section{Introduction}

\parskip 11pt plus 1pt
\setlength{\parindent}{0cm}

It has been demonstrated previously that all the ten-dimensional closed
superstring theories (Type IIA, Type IIB and the two distinct heterotic
superstrings) are hidden in the Hilbert space of the 26-dimensional closed
bosonic string theory
\cite{cent}. More precisely, the states and interaction vertices defining
perturbatively these superstrings in the light-cone gauge form a subset of
states and  interaction vertices of the bosonic string. The emergence of
space-time fermions and of supersymmetry, anticipated by Freund
\cite{freund}, is an impressive property of the bosonic string. The
generation of space-time fermions out of bosons appears in reference
\cite{cent} as a stringy generalisation of the field theoretical mechanism
by which non abelian monopoles become fermions in an appropriate
environment \cite{thooft}.

The superstrings arise from a toroidal compactification of the bosonic
strings on an $E_8\times E_8$ lattice where only the second $E_8$ plays an
active r\^ole. The superstring content of the bosonic string  appears when
all states pertaining to the first $E_8$ are  removed from the spectrum. We
call this
 a ``truncation''. The truncation required to reveal  superstrings has in fact
to be extended to all oscillator states in four of the eight compact
dimensions spanned by the second $E_8$ lattice of zero modes. However,
some particular zero modes  have to be retained in these dimensions.
They  play an essential r\^ole in the construction and have an interesting
interpretation in terms of superghost zero modes.

The  $E_8$ lattice is a sublattice of the
$SO(16)$ lattice and the theory is more elegantly formulated in terms of
$E_8\times SO(16)$ \cite{ens}. This formulation was in fact a crucial step,
because it led to
uncover not only  the  superstrings,  but also the non-supersymmetric
fermionic strings\footnote{We
shall call fermionic strings all strings containing space-time fermions in
separate left or right sectors whether or not they are projected out in the
closed string spectrum.}. Type OA and Type OB \cite{sw} and all
the consistent non-supersymmetric ten-dimensional heterotic strings
discovered in reference
\cite{harveydix} emerged indeed from the
26-dimensional bosonic string compactified on an $E_8\times SO(16)$
lattice, using the same truncation  that generated the superstrings out of its
$E_8\times E_8$ sublattice
\cite{lls,y,ns}.

The universality of the truncation procedure, leading from the bosonic
string to all these fermionic strings,  suggests that this process is a
dynamical one. It was in fact conjectured in reference \cite{ens} that
string field  expectation values in string field theory may provide the
mechanism needed to  move out of the tachyonic vacuum of the bosonic
string to reach the vacuum of a truncated theory. This remained a pure
speculation devoid of any predictive content and there was in fact no clear
indication that truncation had any dynamical significance.

In this paper we extend the truncation procedure to open strings. Our basic
assumption is that the   bosonic string compactified on an $E_8\times
SO(16)$ lattice (or on its  $E_8\times E_8$ sublattice) must be free of
massless tadpole divergences  in leading order in the coupling constant. This
{\em tadpole condition}  is in line with  the philosophy of  building
consistent
``open-descendants'' from the
 closed string sector \cite{sagn1}. There are two types of massless tadpoles
in the  bosonic theory: the dilaton tadpole  and tadpoles associated with
massless scalars originating from the group symmetry of the compactified
theory (hereafter referred to as
${\cal G}$-scalars). Upon truncation all massless modes created by operators
in non-compact dimensions become massive, and a subset  of ${\cal
G}$-scalars are transmuted to new massless modes, including a new dilaton
and a new graviton. The
tadpole condition survives the truncation  for the transmuted modes and we
obtain, {\em from purely bosonic considerations}, the  correct
Chan-Paton groups of the anomaly free Type I theory  and of the
tadpole-free type O theories.

These results are related to the existence
of a {\it unique}    open descendant of the uncompactified
bosonic string satisfying the tadpole condition, namely the unoriented string
with Chan-Paton group
$SO(2^{13})$.  It turns out indeed  that, upon compactification, the rank
of this
Chan-Paton group is reduced exactly to the value required for the fermionic
string generated by  truncation. This effect will be shown  to arise from a
{\it fusion} mechanism whereby many space-filling D-branes coalesce into
{\it one} \footnote{Reduction of the rank of Chan-Paton groups in
compactification has originally been discussed in reference \cite{sagn2}.}.

Interpretation of Chan-Paton multiplicities in terms of D-branes
and orientifold
 leads to yet another prediction. We prove that the
tensions of  fermionic D9-branes arising from truncation equal  the
ten-dimensional tensions of the bosonic wrapped D25-brane and we check
that  the
 correct values  for  the
tensions of  fermionic D9-branes are indeed recovered. This result may be
interpreted as energy conservation  in the truncation process.

The paper is organised as follows. In Section 2, we review in simple terms
all the essential steps needed to understand  how
supersymmetric and non-supersymmetric ten-dimensional fermionic
closed string theories  stem from the compactified closed bosonic string
theory. In Section 3, we consider open string theories. We impose the
tadpole condition on the compactified bosonic string and analyse the fusion
mechanism on Lie algebra lattices. We specialise to
$E_8\times SO(16)$ lattices and  extend the truncation to open string
sectors to get Type I and Type O theories and their D-brane tensions.
In the final section 4, we discuss the significance of our results and stress
their possible relevance  for a fundamental theory of quantum gravity.

\section{Truncation of the closed bosonic  string}

\subsection{Fermions and world-sheet supersymmetry}

To accommodate space-time fermions   in
 the  26-dimensional bosonic string one must meet three requirements:

\hskip 1cm a) A continuum of bosonic zero modes must be removed. This
can be achieved by compactifying  $d=24-s$ transverse dimensions  on a
$d$-dimensional torus. This leaves $s+2$ non-compact dimensions with
transverse group
$SO_{trans}(s)$.

\hskip 1cm b) Compactification must generate an internal group $SO_{int}
(s)$ admitting spinor representations\footnote{We shall  designate all
locally isomorphic groups by the same notation.}. This can be achieved by
toroidal compactification on the Lie lattice of a simply laced Lie group
${\cal G}$ of rank $d$ containing a subgroup $SO_{int} (s)$. The latter is
then mapped onto  $SO_{trans} (s)$  in such a way that the diagonal algebra
$ so_{diag}(s) ={\rm diag} [so_{trans}(s) \times so_{int}(s)]$ becomes
identified with a new transverse algebra. In this way, the spinor
representations of  $SO_{int} (s)$ describe  fermionic states because a
rotation in space induces a half-angle rotation on these states.  This
mechanism is  distinct from the two-dimensional world-sheet equivalence of
bosons and fermions. It is reminiscent of a similar mechanism at work in
monopole theory: there, the diagonal subgroup of space-time rotations and
isospin rotations can generate space-time fermions from a bosonic field
condensate in spinor representations of the isospin group~\cite{thooft}.

\hskip 1cm c)  The consistency of the above procedure relies on the
possibility of extending the diagonal algebra $so_{diag}(s)$ to the new full
Lorentz algebra
$so_{diag}(s+1,1)$,  a highly non trivial constraint. To break the original
Lorentz group $SO(25,1)$ in favour of the new one,   a truncation consistent
with conformal invariance must be performed on the physical spectrum of
the bosonic string. Actually, states described by 12 compactified bosonic
fields must be truncated, except possibly for zero modes. This  follows
from the need, in string theory, of a local world-sheet supersymmetry  to
accommodate space-time fermions. In units where each two-dimensional
boson contributes $1$ to the Virasoro central  charge, the related
superghost contributes $11$, while the longitudinal and time-like Majorana
fermions contribute $2 \times 1/2$.  Therefore, one requires
$12$ compactified bosons to account for the anomaly of the superghost and
unphysical Majorana fermions.   Moreover, the need to generate an internal
group $SO_{int}(s)$ via toroidal compactification requires $s/2$
compactified bosons which can account for $s$ transverse Majorana
fermions (we hereafter take $s$ to be even, in which case $s/2$ is the rank
of the internal group). Therefore, one must ensure that the total number
$d=24-s$ of compactified dimensions is at least $12 + s/2$. In other words,
\begin{equation} \label{dimension} s\leq 8\ ,
\end{equation} and the     highest available space-time dimension
accommodating fermions is therefore
$s+2=10$ \cite{cent, ens}.

 We shall verify below that the closure of the Lorentz algebra indeed
requires Eq.(\ref{dimension}).
\subsection{Closed string compactification and zero modes}
 Consider the bosonic closed string compactified on a
$d$-dimensional torus. In terms of the left and right compactified momenta,
the mass spectrum is
\begin{eqnarray}
\label{spectrum} {\alpha^\prime m_L^2\over 4}&=&  \alpha^\prime {\bf
p}_L^2 + N_L -1\ , \nonumber \\ {\alpha^\prime m_R^2\over 4}&=&
\alpha^\prime {\bf p}_R^2 + N_R -1 \ , \\ &\hbox{and}&\nonumber\\ m^2 = {
m_L^2\over 2}+{ m_R^2\over 2}\quad &;&\quad m_L^2=m_R^2\ .
\end{eqnarray} In Eq.(\ref{spectrum})
$N_L$ and
$N_R$ are the oscillator numbers in 26-dimensions and the zero modes
$\sqrt{2\alpha^\prime}{\bf p}_L$,
$\sqrt{2\alpha^\prime}{\bf p}_R$ span a
$2d$-dimensional even self-dual Lorentzian lattice with negative (resp.
positive) signature for left (resp. right) momenta. This ensures modular
invariance of the closed string spectrum  \cite{narain}. For generic toroidal
compactifications, the massless vectors
$\alpha_{-1,R}^\mu ~\alpha_{-1,L}^i\vert 0_L,0_R\rangle$ and
$\alpha_{-1,L}^\mu ~\alpha_{-1,R}^i \vert 0_L,0_R\rangle$, where the
indices
$\mu$ and
$i$ respectively refer to non-compact and  compact dimensions, generate a
local symmetry  $[U_L(1)]^d\times [U_R(1)]^d$. But more massless vectors
arise when $ \sqrt{2\alpha^\prime}{\bf p}_L$ and
$\sqrt{2\alpha^\prime}{\bf p}_R$ are  roots of simply laced groups
${\cal G}_L$ and ${\cal G}_R$ (with roots length $\sqrt{2}$). The gauge
symmetry is  enlarged to  ${\cal G}_L\times {\cal G}_R$  and the
theory is modular invariant provided the lattice of zero modes  is  self-dual
Lorentzian even. In particular, this is the case for compactification on  a
${\cal G}\times {\cal G}$ lattice where
${\cal G}$ is any semi-simple simply laced group  of rank $d$, if both
$\sqrt{2\alpha^\prime}{\bf p}_L$ and $\sqrt{2\alpha^\prime}{\bf p}_R$
span the full weight lattice
$\Lambda_{weight}$ of
${\cal G}$,  but are constrained to be in the same conjugacy class. Namely
 $\sqrt{2\alpha^\prime}({\bf p}_L-{\bf p}_R)$ must be on the root lattice
$\Lambda_{root}$ of ${\cal G}$ \cite{en}.  Hereafter, such lattice will be
referred  to as the EN lattice of ${\cal G}$ .

Now perform  a toroidal compactification on  a ${\cal G}_L\times {\cal G}_R$
lattice. Then truncate the spectrum of, say, the right sector of the theory, by
tentatively removing {\em all} states (i.e. oscillators and momenta) in $24-
3s/2$  of the total number $d=24-s$ of compact dimensions, leaving $s+2$
non-compact dimensions. Also, since the truncation  removes a  subset of
states whose compactified momenta lie on the weight lattice of ${\cal
G}_R$, one expects the embedding of the subgroup
$ SO_{int}(s)$ of ${\cal
G}_R$ to be  regular\footnote{An embedding is regular if the root lattice of
the subgroup is  contained in the root lattice of the group.}.
This must in fact be the case as discussed below.

Let us examine the closure of the Lorentz algebra $ so(s+1,1)$ \cite{cent,
ens}. The generators of the diagonal  group $ SO_{diag}(s)$ are given by
\begin{equation}
\label{diagonal} J^{ij} = L^{ij} + K_o^{ij}\ ,~~~~~~~i,j~=~1,...,s \ ,
\end{equation} where $L^{ij}$ are the  generators of the transverse
space-time group
$SO_{trans}(s)$ and the operators $K_o^{ij}$ generate the internal
$SO_{int}(s)$. Note that Eq.(\ref{diagonal}) maps the internal group onto
$SO_{trans}(s)$. The
$K_o^{ij}$  are the zero modes of the full Kac-Moody algebra generated in
the compact dimensions from   bosonic vertex operators  and obey the
commutation relations
\begin{equation}
\label{KM} [ K_m^{ij} , K_n^{lp}] = i (K_{m+n}^{ip}\, \delta ^{jl} +
K_{m+n}^{jl}\, \delta ^{ip} -K_{m+n}^{il}\, \delta ^{jp}- K_{m+n}^{jp}\,
\delta ^{il}) +2mk
\delta_{pl} ^{ij}\delta_{m+n,o}
\end{equation} where
\begin{equation}
\delta_{pl} ^{ij}={1\over2} (\delta_p^i \delta_l^j - \delta_l^i \delta_p^j)\ ,
\end{equation} and $k$ is the level of the algebra. To check the closure of
the Lorentz algebra, one has to supplement Eq.(\ref{diagonal}) with the
generators
$J^{i+}, J^{+-}$ and $J^{i-}$ defined in the light-cone gauge. The first $s+1$
operators only involve position and momentum operators, but the
$J^{i-}$  involve oscillators in $\nu=s/2+s$ compact and transverse
non-compact dimensions. The Kac-Moody generators Eq.(\ref{KM})
are needed in the $J^{i-}$ to ensure their correct commutation relations with
the $J^{ij}$. It remains to show that boosts commute, i.e.
\begin{equation}\label{boosts}
\label {commute} [J^{i-}, J^{j-}] =0 \ .
\end{equation}  This is a non-trivial requirement. Imposing the mass shell
condition
\begin{equation}
\label{mass} {\alpha^\prime m_R^2\over 4}= \alpha^\prime {\bf
p}_{R}^2[SO(s)] + N_R(\nu) -C \ ,
\end{equation}  where  $N_R(\nu)$ is the oscillator number in $\nu$
dimensions and
 $C$ is a parameter, one obtains Eq.(\ref{boosts}) provided  \cite{ens}
\begin{equation} {\nu\over 12} = 2-k=2C\ .
\end{equation}  The solution\footnote {The solution
$k=0, \nu=24, C=1$ has nothing to do with  an internal group $SO_{int}(s)$.
There is no contribution from a Kac-Moody algebra;  we simply recover the
original bosonic string Lorentz group for any compactification and the
constant $C=1$ agrees with the mass spectrum Eq.(\ref{spectrum}). The
case
$k=2, \nu=C=0$ will not be discussed here. }  $ \nu=12, k=1$, $C=1/2$,  is
compatible with our truncation scheme. Recall indeed that $\nu$ counts the
number of compact dimensions associated with
$SO_{int}(s)$ as well as the $s$ transverse space-time dimensions. So,
$\nu=12$ corresponds to $s=8$ and,  in accordance with the central charge
argument developed in section {\bf 2.1}, to a truncation of twelve
compactified dimensions.\footnote{The above derivation of the closure
applies trivially when some additional dimensions of the ten-dimensional
strings  are compactified.} Furthermore, compactifications on lattices of
simply laced groups always generate level one  Kac-Moody algebras
\cite{halpern}, and  therefore,  $k=1$ is consistent with  torus
compactification of $SO_{int}(s)$. Moreover, the Dynkin index of its
embedding in  ${\cal G}_R$ is  $k[
\widehat{SO_{int}(s)}]/ k[\widehat{{\cal G}_R }]$ \cite{betz}. This is equal to
one, which implies that the embedding is regular.

There is however a potential problem of compatibility between the closure
of the new Lorentz algebra and our truncation: setting $C=1/2$ in
Eq.(\ref{mass}) is not consistent with Eq.(\ref{spectrum}) if the truncation
only keeps the zero mode contribution coming from the internal symmetry
group $SO_{int}(8)$. There is  some subtlety with zero modes in the
other compact dimensions. To clarify the issue, let us further analyse the
embedding of $SO_{int}(8)$ in
${\cal G}_R$ and the twelve-dimensional truncation.

The centre of the covering group of $SO(8)$ is  $Z_2\times Z_2$. Its   four
elements partition the weight lattice in  four conjugacy classes
$(o)_8,(v)_8, (s)_8, (c)_8$  isomorphic to the root lattice. The $(o)_8$
lattice is the root lattice itself and contains the element  $
\sqrt{2\alpha^\prime}{\bf p_o}=(0,0,0,0)$. The
$(v)_8$ lattice is the vector lattice whose smallest weights are eight
vectors of norm one; in an orthonormal basis, these are
$\sqrt{2\alpha^\prime}{\bf p_v}=(\pm1, 0,0,0)$ + permutations. The $(s)_8$
and $(c)_8$ lattices are spinor lattices whose smallest weights also have
norm one and are the eightfold degenerate vectors
$\sqrt{2\alpha^\prime}{\bf p_{s,c}}=(\pm1/2,\pm1/2,\pm1/2,\pm1/2)$
with even (for class $(s)_8$) or odd (for class $ (c)_8$) number of minus
signs. The structure of the weight lattice of all $SO(4m)$ groups is the
same: in a
$2m$-dimensional Cartesian basis, the root lattice vectors have integer
components whose sum is even (and contains the element
$\sqrt{2\alpha^\prime}{\bf p_o}={\bf 0}$). The vector
$\sqrt{2\alpha^\prime}{\bf p_v}=(\pm1, 0 .....)$ still has norm one but the
spinors $\sqrt{2\alpha^\prime}{\bf p_{s,c}}=(\pm1/2, \pm1/2.....)$ have
norm increasing with $m$. The degeneracy in norm of $(v)_8, (s)_8$ and
$(c)_8$ in
$SO(8)$ is rooted in the triality properties of the group, and the choice of a
vector representation $\sqrt{2\alpha^\prime}{\bf p_v}$  is a mere
convention.
$\sqrt{2\alpha^\prime}{\bf p_v}$ is in fact defined by its mapping onto the
representation of the
$SO_{trans}(8)$ group as described in Eq.(\ref{diagonal}). It is this mapping
which transmutes the spinors $(s)_8$ and $(c)_8$ of $SO_{int}(8)$ to
space-time spinors of the Lorentz group $SO(9,1)$.

In order for Eq.(\ref{spectrum}) to be compatible with Eq.(\ref{mass}) one
must keep zero modes in the 16 compact dimensions  in such a way that
\begin{equation}
\label{ghost} \alpha^\prime{\bf p}_R^2[{\cal G}_R] = \alpha^\prime{\bf
p}_R^2[SO(8)] +{1\over 2}\ .
\end{equation} Note that the contribution  of these zero modes to the
spectrum, namely $1/2$, is exactly the energy required to remove the
zero-point energy (-12/24) contribution to the energy of the states taken
out by the truncation.

As shown below, Eq.(\ref{ghost}) can be satisfied by choosing
\begin{equation}
\label{group} {\cal G}_R= E_8 \times SO(16)\ ,
\end{equation} and we shall prove at the end of section {\bf 2.4} that from
this choice  all  known ten-dimensional closed fermionic strings
emerge from the same truncation scheme.  We
decompose
$SO(16)$  in $SO^{\,
\prime}(8) \times SO(8)$ and truncate all  states created  by oscillators in
the 12 dimensions defined by the $E_8\times SO^{\,\prime}(8)$ root lattice.
To satisfy  Eq.(\ref{ghost}), we keep only  $SO(16)$ zero modes which are
$SO^{\,\prime}(8)$ vectors of norm one. The latter are  chosen as follows.

The decomposition of an $SO(16)$ lattice in terms of
$SO^{\,
\prime}(8)
\times SO(8)$ lattices yields
\begin{eqnarray} (o)_{16} = [(o)_{8^\prime}\oplus (o)_{8}] &+&
[(v)_{8^\prime}\oplus (v)_{8}] \ ,\nonumber\\ (v)_{16} =
[(v)_{8^\prime}\oplus (o)_{8}] &+& [(o)_{8^\prime}\oplus (v)_{8}]\nonumber\
,\\ (s)_{16} = [(s)_{8^\prime}\oplus (s)_{8}]& +&[ (c)_{8^\prime}\oplus
(c)_{8}]\nonumber\ ,\\ (c)_{16} = [(s)_{8^\prime}\oplus (c)_{8}] &+&
[(c)_{8^\prime}\oplus (s)_{8}]\ .
\end{eqnarray}
The vectors of norm one in $SO^{\, \prime}(8)$ are the
4-vectors $\sqrt{2\alpha^\prime}{\bf p^\prime_v},
\sqrt{2\alpha^\prime}{\bf p^\prime_s}$ and
$\sqrt{2\alpha^\prime}{\bf p^\prime_c}$ described above. We choose one
vector
$\sqrt{2\alpha^\prime}{\bf p^\prime_v}$ and one vector
$\sqrt{2\alpha^\prime}{\bf p^\prime_s}$ (or equivalently
$\sqrt{2\alpha^\prime}{\bf p^\prime_c}$).

This gives the truncations
\begin{eqnarray}
\label{truncations} &&(o)_{16} \rightarrow   (v)_{8} \qquad (v)_{16}
\rightarrow   (o)_{8}\ ,\nonumber\\ &&(s)_{16} \rightarrow   (s)_{8} \qquad
(c)_{16} \rightarrow   (c)_{8} \ ,\\
\hbox{or}\nonumber \\ \label{truncationc} &&(o)_{16} \rightarrow   (v)_{8}
\qquad (v)_{16}
\rightarrow   (o)_{8}\ ,\nonumber\\ &&(s)_{16} \rightarrow   (c)_{8} \qquad
(c)_{16} \rightarrow   (s)_{8} \ .
\end{eqnarray}
It follows from the closure of the Lorentz algebra that
states belonging to the lattices $(v)_{8}$ or $(o)_{8}$   are bosons while
those belonging to the spinor lattices $(s)_{8}$ and $(c)_{8}$ are
space-time fermions. These zero modes ensure the truncation consistency
by selecting, in the light-cone gauge, the emission vertices of the fermionic
strings as subsets of the emission vertices of the bosonic string \cite{cent,
ens}. They may in fact be viewed as     superghosts zero modes
entering  emission vertices  in the fermionic string\footnote{For a
general discussion on the mapping of the states of the truncated bosonic
string to the covariant   bosonic formulation of fermionic strings, see
reference
\cite{revlsw}.}.  We shall therefore refer to these zero modes as to ghost
vectors.  We now show that this choice of zero modes
$\sqrt{2\alpha^\prime}{\bf p^\prime_v}$ and
$\sqrt{2\alpha^\prime}{\bf p^\prime_s}$ (or $\sqrt{2\alpha^\prime}{\bf
p^\prime_c}$)  preserve modular invariance in the truncation.

\subsection{Modular invariance}

Consider a toroidal compactification of closed strings on a
${\cal G}_L\times{\cal G}_R$ lattice which is Lorentzian self-dual even.
 Instead of starting with the modular invariant
partition function of this lattice, we shall analyse the modular
transformation  properties of the partition function of the full weight lattice
$\Lambda^*
\equiv\Lambda ({\cal G}_L\times{\cal G}_R)$. This lattice admits a coset
decomposition
$\Lambda^*/ \Lambda_{11}$ with  $\Lambda_{11}$  the even (but in general
 not self-dual) Lorentzian lattice of the vectors
$\sqrt{2\alpha^\prime}{\bf p}_{11} = (\sqrt{2\alpha^\prime}{\bf p}_
{1L},\sqrt{2\alpha^\prime} {\bf p}_{1R})$, where
$\sqrt{2\alpha^\prime}{\bf p}_ {1L}$ and
$\sqrt{2\alpha^\prime}{\bf p}_ {1R}$ span the root lattices of
${\cal G}_L$ and ${\cal G}_R$. The coset decomposition follows from the
fact that any even lattice is integral (i.e.  for all pairs  of lattice vectors
${\bf v}$ and ${\bf v^\prime}$, ${\bf v}.{\bf v^\prime}\in Z$) and that any
integral lattice is a
 sublattice of its dual. The latter can then be decomposed into  an integer
number
${\cal N}$ of cosets. In the present case, we have
\begin{equation}
\label{coset} \Lambda^* =  \bigoplus_{\alpha=1}^{{\cal
N}_L}
\bigoplus_{\beta=1}^{{\cal N}_R}\Lambda_{\alpha \beta}\ ,
\end{equation} where the ${\cal N}={\cal N}_L{\cal N}_R$ sublattices
$\Lambda_{\alpha \beta}$ are isomorphic to the lattice $\Lambda_{11}$.
${\cal N}_L$ and ${\cal N}_R$ are equal to the order of the centres  of the
covering groups of ${\cal G}_L$ and
${\cal G}_R$.

The partition function of each sublattice $\Lambda_{\alpha \beta}$
factorises in left and right partition functions
\begin{equation}
\label{factor}
\gamma_{\alpha\beta} (\bar\tau,\tau)={\bar \gamma}_{\alpha
L}(\bar\tau)\gamma_{\beta R}(\tau)\ ,
\end{equation} where
\begin{equation}
\label{partition}
\gamma_{\beta R} (\tau)=\sum_{\sqrt{2\alpha^\prime}{\bf p}_{1R}\in
\Lambda_{1R}} \exp
\{2\pi i\tau[ \alpha^\prime({\bf p}_{1 R}+{\bf p}_{\beta R} )^2 + N_R^{(c)}
-{\delta_R \over 24 }]\}\ .
\end{equation} Here $\Lambda_{1R}$ is the ${\cal G}_R$ root lattice and
${\bf p}_{\beta R}$ is an arbitrarily chosen vector of a sublattice
$\Lambda_{\beta R}$ in the coset decomposition $\Lambda_{1R}^*/
\Lambda_{1R}$,  where $\Lambda_{1R}^*$ is the
${\cal G}_R$~weight lattice. $ N_R^{(c)}$ is the oscillator number in the
compact dimensions and  $\delta_R$ is the number of compact
dimensions. A similar expression holds for ${\bar \gamma}_{\alpha
L}(\bar\tau)$. Clearly, if ${\cal G}_R$  can be further decomposed  in a
direct product of simply laced groups, $\gamma_{\beta R}(\tau)$ can be
further factorised  accordingly. The  $S$ modular transformation ($S :
\tau \to -1/\tau$) of each factor\footnote{This expression is a
straightforward consequence of the Poisson resummation formula and of
the relations  $V(\Lambda_{11})=\sqrt {\cal N}$  ($V(\Lambda^*)=1/ \sqrt
{\cal N}$).  See for instance  reference \cite{revlsw}.} (e.g.)
\begin{equation}
\label{inversion}
\gamma_{\beta R}(\tau)={1\over \sqrt {{\cal N}_R} }\sum_{\eta =1}^{{\cal
N}_R}\exp (4\pi i \alpha^\prime{\bf p}_{\beta R}.{\bf p}_{\eta R})\
 \gamma_\eta (-{1\over \tau})\ ,
\end{equation} provides a simple proof that {\em any modular invariant
closed bosonic string}, compactified on a
${\cal G}_L\times{\cal G}_R$  lattice,  (with ${\cal G}_R$ and/or  ${\cal
G}_L$ =
$E_8\times SO(16))$, {\em yields after truncation a
fermionic modular invariant ten-dimensional string.}

 Eq.(\ref{inversion}) shows that under the $S$ modular
transformation the  partition functions of the four
 sublattices $(o)_{16},(v)_{16},(s)_{16},(c)_{16}$ transform as
\begin{eqnarray}
\label{sixteen}
\gamma_{(o)_{16}} &\rightarrow& {1\over2} [
\gamma_{(o)_{16}}+\gamma_{(v)_{16}}+\gamma_{(s)_{16}}+\gamma_{(c)_{16}}]
\ ,\nonumber\\
\gamma_{(v)_{16}} &\rightarrow& {1\over2} [
\gamma_{(o)_{16}}+\gamma_{(v)_{16}}-\gamma_{(s)_{16}}-\gamma_{(c)_{16}}]
\ ,\nonumber\\
\gamma_{(s)_{16}} &\rightarrow& {1\over2} [
\gamma_{(o)_{16}}-\gamma_{(v)_{16}}+\gamma_{(s)_{16}}-\gamma_{(c)_{16}}]
\ ,\nonumber\\
\gamma_{(c)_{16}} &\rightarrow& {1\over2} [
\gamma_{(o)_{16}}-\gamma_{(v)_{16}}-\gamma_{(s)_{16}}+\gamma_{(c)_{16}}]
\ ,
\end{eqnarray} as do the four $SO(8)$ partition functions
$\gamma_{(o)_{8}},\gamma_{(v)_{8}},\gamma_{(s)_{8}}$ and
$\gamma_{(c)_{8}}$. The crucial point is that our truncation
Eqs.(\ref{truncations}) or (\ref{truncationc}), together with a sign flip on
the spinorial $SO(8)$ partition functions
$\gamma_{(s)_{8}}$ and
$\gamma_{(c)_{8}}$,  commutes with $S$. We show this for the truncation
with ghost vector $\sqrt{2\alpha^\prime}{\bf p^\prime_s}$ (the proof is
similar for $\sqrt{2\alpha^\prime}{\bf p^\prime_c}$). Truncating and
flipping
$\gamma_{(o)_{16}}$ as well as its $S$-transform as given in
Eq.(\ref{sixteen}), we get the correct $S$-transform of
$\gamma_{(v)_{8}}$, as depicted in the diagram below.
\begin{center}
\begin{tabular}{ccc}
$\gamma_{(o)_{16}}$ &$\longrightarrow $& ${1\over2} [
\gamma_{(o)_{16}}+\gamma_{(v)_{16}}+\gamma_{(s)_{16}}+\gamma_{(c)_{16}}]
$\\ &$S$&\\ &&\\ Trunc $\downarrow $
Flip\hbox{\phantom{z}\phantom{z}}  & & Trunc $\downarrow $
Flip\hbox{\phantom{z}\phantom{z}} \\ &&\\&$S$&\\
$\gamma_{(v)_{8}}$&$\longrightarrow$ & ${1\over2} [
\gamma_{(v)_{8}}+\gamma_{(o)_{8}}-\gamma_{(s)_{8}}-\gamma_{(c)_{8}}]$
\end{tabular}
\end {center} Similarly, starting with $\gamma_{(s)_{16}}$, we get the
corresponding diagram.
\begin{center}
\begin{tabular}{ccc}
$\gamma_{(s)_{16}}$ &$\longrightarrow $& ${1\over2} [
\gamma_{(o)_{16}}-\gamma_{(v)_{16}}+\gamma_{(s)_{16}}-\gamma_{(c)_{16}}]
$\\ &$S$&\\ &&\\ Trunc $\downarrow $
Flip\hbox{\phantom{z}\phantom{z}}  & & Trunc $\downarrow $
Flip\hbox{\phantom{z}\phantom{z}} \\ &&\\&$S$&\\
$-\gamma_{(s)_{8}}$&$\longrightarrow$ & ${1\over2} [
\gamma_{(v)_{8}}-\gamma_{(o)_{8}}-\gamma_{(s)_{8}}+\gamma_{(c)_{8}}]$
\end{tabular}
\end {center}
 The commutation with $S$ of the truncation-flip for the two
remaining partition functions follows from the above diagrams in an
obvious way. It follows from the commutation properties just described
and from the factorisation  of the  partition function of the lattice
$\Lambda^* \equiv\Lambda ({\cal G}_L\times{\cal G}_R)$, Eq.(\ref{factor}),
that any $S$~invariant partition function before truncation and flip
remains so afterwards.

To summarise, the truncated partition functions do  have the same modular
properties under inversion $S$ as the original bosonic partition functions
{\em provided, in accordance with the $SO(9,1)$ Lorentz group, internal
spinors are transmuted to space-time spinors}. A glance at
Eq.(\ref{partition}) shows that the invariance under the translations
$\tau
\rightarrow \tau +1$,
$\bar\tau
\rightarrow \bar\tau +1$ is also preserved when both sectors are truncated.
It is preserved for the heterotic strings  because the half integer
$(\delta_R - \delta_L)/24 = (4-16)/24 $  compensates the half integer
arising from  ghosts of norm one. The  modular invariance of the lattice
partition function ensures the modular invariance of the partition function
of the full theory. This completes the proof that the truncation from
$E_8\times SO(16)$ to $ SO_{int}(8) + ghosts$ transfers modular
invariance from the 26-dimensional bosonic string to  ten-dimensional
fermionic strings\footnote{For an alternate proof, see ref. \cite{ns}.
Generalisation to multiloops is given in ref. \cite{s}.}.

\subsection{Closed fermionic strings}

We now explain how the consistent fermionic ten-dimensional strings
\cite{sw, harveydix} are   obtained by truncation from the 26-dimensional
bosonic string.

First we consider the fermionic theory emerging from truncation in both
left and right sectors. We thus examine a compactification on both sectors
with
${\cal G}_L={\cal G}_R=E_8\times SO(16)$. In the truncated theory,
$(E_8)_{L,R}$ merely disappear, and we therefore only discuss in detail the
fate of the $SO(16)_{L,R}$ representations under truncation. In the full
bosonic string,  the representations of
$SO(16)$ entering the partition function are restricted by modular
invariance\footnote{Strictly speaking the modular invariance does not fully
determine the representations of $SO(16)$ at this level (because
$\gamma(s)=\gamma(c)$). However at the level of the amplitudes, this
ambiguity is lifted.}.  Typically, such partition function is a sum of
products of left and right partition functions
$\bar\gamma_{\alpha L},\gamma_{\alpha R}$.

Using the transformation properties Eq.(\ref{sixteen}), one finds that there
are two distinct modular invariant partition functions. The first one is,
\begin{eqnarray}
\label{first} (\bar\gamma_{(o)_{16},L}+\bar\gamma_{(s)_{16},L})
(\gamma_{(o)_{16},R}+\gamma_{(s)_{16},R})\nonumber\\ &&\hskip
-8cm=\bar\gamma_{(o)_{16},L}\gamma_{(o)_{16},R}+
\bar\gamma_{(s)_{16},L}\gamma_{(o)_{16},R} +\bar\gamma_{(o)_{16},L}
\gamma_{(s)_{16},R}+\bar\gamma_{(s)_{16},L}\gamma_{(s)_{16},R}\ ,
\end{eqnarray} and the second is,
\begin{equation}
\label{second}
\bar\gamma_{(o)_{16},L}\gamma_{(o)_{16},R}+\bar\gamma_{(v)_{16},L}
\gamma_{(v)_{16},R}+
\bar\gamma_{(s)_{16},L}\gamma_{(s)_{16},R}+\bar\gamma_{(c)_{16},L}\gamma_{(c)_{1
6},R}\ .
\end{equation}
 Eq.(\ref{first}) can be rewritten as
$\bar\gamma_{(o)_{E8},L}\gamma_{(o)_{E8},R} $ where the subscript
$(o)$ refers to the  root lattice of  $E_8$ which is a sublattice of the
$SO(16)$ weight lattice.  Eqs.(\ref{first}) and (\ref{second}) describe
compactifications on the EN lattices  of   $ E_8\times E_8$  and of  $
E_8\times SO(16)$.

 To interpret the result of the truncation in conventional terms, we note that
the $SO(8)$ partition  functions
$\gamma_{(o)_{8}}$ and $\gamma_{(v)_{8}}$, divided by the Dedekind
functions arising from the bosonic states in the eight  non
compact transverse dimensions are  the Neveu-Schwarz partition functions
with the `wrong' and `right' GSO projection $(NS)_-$ and
$(NS)_+$.  The partition  functions
$\gamma_{(s)_8}$ and $\gamma_{(c)_8}$, divided  by the same Dedekind
functions,  form the two Ramond partition functions of opposite chirality
$R_+$ and
$R_-$.

Choosing the ghosts
$\sqrt{2\alpha^\prime}{\bf p^\prime_v}$  and
$\sqrt{2\alpha^\prime}{\bf p^\prime_s}$ in both sector, and truncating in
accordance with Eq.(\ref{truncations}), the first partition function
Eq.(\ref{first}) yields the supersymmetric chiral closed string,
\begin{center}
$\bar\gamma_{(o)_{E8},L}\gamma_{(o)_{E8},R}
= \bar\gamma_{(o)_{16},L}\gamma_{(o)_{16},R}+
\bar\gamma_{(s)_{16},L}\gamma_{(o)_{16},R} +\bar\gamma_{(o)_{16},L}
\gamma_{(s)_{16},R}+\bar\gamma_{(s)_{16},L}\gamma_{(s)_{16},R}$

Trunc $\downarrow$ Flip

$\bar\gamma_{(v)_{8},L}\gamma_{(v)_{8},R}-
\bar\gamma_{(s)_{8},L}\gamma_{(v)_{8},R} -\bar\gamma_{(v)_{8},L}
\gamma_{(s)_{8},R}+\bar\gamma_{(s)_{8},L}\gamma_{(s)_{8},R}$\ .
\end {center}
Taking into account the oscillators in non compact dimensions, this is
\begin{equation}
\label{twob}
 {\bf IIB}: \quad (NS)_+\ (NS)_+ + R_+ \ (NS)_+ + (NS)_+ \ R_+ + R_+\ R_+ \ .
\end{equation}
Replacing  ${\bf p^\prime_s}$ by  ${\bf p^\prime_c}$ in, say, the right
sector we get, using Eq.(\ref{truncationc}), the non chiral supersymmetric
closed string
\begin{center}
$\bar\gamma_{(o)_{E8},L}\gamma_{(o)_{E8},R}=\bar\gamma_{(o)_{16},L}\gamma_{(o)_{
16},R}+
\bar\gamma_{(s)_{16},L}\gamma_{(o)_{16},R} +\bar\gamma_{(o)_{16},L}
\gamma_{(s)_{16},R}+\bar\gamma_{(s)_{16},L}\gamma_{(s)_{16},R}$

Trunc $\downarrow$ Flip

$\bar\gamma_{(v)_{8},L}\gamma_{(v)_{8},R}-
\bar\gamma_{(s)_{8},L}\gamma_{(v)_{8},R} -\bar\gamma_{(v)_{8},L}
\gamma_{(c)_{8},R}+\bar\gamma_{(s)_{8},L}\gamma_{(c)_{8},R}\ .$
\end {center}
Taking into account the oscillators in non compact dimensions, this is
\begin{equation}
\label{twoa}
 {\bf IIA}: \quad (NS)_+\ (NS)_+ + R_+ \ (NS)_+ + (NS)_+ \ R_- + R_+\ R_-\ .
\end{equation}
The same choices of the ghosts in the second partition function yield the
non-supersymmetric strings
\begin{equation}
\label{ob}
 {\bf 0B}: \quad (NS)_+\ (NS)_+ +(NS)_-\ (NS)_- + R_+\ R_++ R_- \ R_- \ ,
\end{equation}
\begin{equation}
\label{oa}
 {\bf 0A}: \quad (NS)_+\ (NS)_+ +(NS)_-\ (NS)_- + R_+\ R_-+ R_- \ R_+ \ .
\end{equation}

 Heterotic strings are generically obtained from compactification on
${\cal G}_L\times {\cal G}_R$ by only truncating in the right sector with
${\cal G}_R= E_8 \times SO(16)$. In the general case, the partition function
constructed on  ${\cal G}_L\times [E_8 \times SO(16)]_R$ must  be modular
invariant. Because  of the factorization Eq.(\ref{factor}) we may replace
the Lorentzian metric by a Euclidean one and drop the $E_8$ to preserve
invariance under translation $(\tau \to \tau +1)$ in the Euclidean metric.
This reduces the problem of finding all heterotic strings obtainable in this
way  to that of finding all 24-dimensional Euclidean even self-dual lattices
containing a sublattice
$\Lambda(SO(16))$. All 24-dimensional even self-dual  Euclidean  lattices
have been classified: they are known as the  Niemeier lattices. Heterotic
strings are obtained from the relevant Niemeier lattices by the truncation
Eq.(\ref{truncations}) (or equivalently, by Eq.(\ref{truncationc})) \cite{lls}.
In particular, one recovers the usual supersymmetric heterotic strings
when  only the sublattices
$(o)_{16}~\oplus~(s)_{16}=~\Lambda (E_8)$ are kept in $\Lambda(SO(16))$.
Indeed, this is an even  self-dual  Euclidean lattice, and therefore, the
lattice
$\Lambda ( {\cal G}_L)$ must also be even self-dual Euclidean, namely
$\Lambda(E_8 \times E_8)$ or
$\Lambda(Spin(32)/Z_2)=(o)_{32} \oplus (s)_{32}$\footnote{All these
heterotic strings have rank 16 groups ${\cal G}_L$. There is one heterotic
string with left symmetry $E_8$. Although it can still be derived by the
general truncation procedure from the bosonic string, it involves an
additional twist
\cite {y}.}.

\section{ Brane fusion and the open string sectors}

We  now turn to the 10-dimensional open string theories:  the
supersymmetric Type I theory and the non-supersymmetric Type O
theories.

We shall impose the {\em tadpole condition} on the bosonic string theory,
namely we impose that  massless tadpoles do not contribute to vacuum
amplitudes in leading order in the coupling constant.

After reviewing the well-known  result
that, for the uncompactified bosonic string, this condition leads to the
unique unoriented string theory with Chan-Paton group $SO(2^{13})$
\cite{sob}, we  study its implications  for strings compactified on Lie
algebra lattices. We  specialise to  $E_8\times E_8$  and  $E_8\times
SO(16)$ lattices,  and  perform the truncation to obtain the  Type I theory
and  the Type O tadpole-free theories.

We interpret the reduction of the Chan-Paton multiplicities which led to
these results in terms  of  orientifolds and D-branes. We make explicit the
{\it fusion} mechanism whereby many space-filling D-branes coalesce into
{\it one}.    Fusion  relates all fermionic open string theories to the
ancestor
bosonic $SO(2^{13})$ theory.

Finally we  prove that energy is conserved in the truncation process.

\subsection{The bosonic open string ancestor}

We review the derivation of the existence of  a
26-dimensional   open bosonic string  free of massless tadpole
divergences\footnote{In this paper, the elimination of massless
tadpole divergences is always only imposed in leading order in the coupling
constant.}.

The Chan-Paton group of the 26-dimensional unoriented, uncompactified, open
string theory may be fully determined by the  tadpole
condition \cite{sagn1,cal,cai}. The full one-loop vacuum amplitude  of a theory
with open and closed unoriented strings comprises the four loop amplitudes
with vanishing Euler characteristic: the torus ${\cal T }$, the Klein bottle
${\cal K }$, the annulus ${\cal A}$ and the M{\"o}bius strip ${\cal M}$. The
last three amplitudes contain ultraviolet divergences which are
conveniently analysed in  the transverse channel.  This  channel describes
the tree level exchange  of zero momentum closed string
modes between holes and/or crosscaps.   The divergences appear there in
the infrared limit and  are associated with the exchange of tachyonic and
massless modes.  One ensures the  tadpole condition, that is the
cancellation of the  divergences due to the massless modes in the total
amplitude, by fixing the Chan-Paton group. Here, the
divergence is related  to  the  dilaton tadpole,
a non-zero one point function of a closed vertex operator
on the disk or on the projective plane.  If the tadpole
condition is not imposed,  the low-energy  effective action acquires a
dilaton potential. Let us stress that in the present case  the tadpole
condition  defining the bosonic open string    is not
compulsory: the presence of the dilaton
tadpole does not  render the  theory  inconsistent if the vacuum is shifted
by the Fishler-Susskind mechanism \cite{fish}.

The tadpole condition for the 26-dimensional uncompactified bosonic string
determines the Chan-Paton group to be $SO(2^{13})$. Let us
discuss  the derivation of this well-known result. Introducing a Chan-Paton
multiplicity
$n$ at  both string ends, the four  different one-loop amplitudes
of the unoriented 26-dimensional bosonic string  are given  by
(see for example \cite{pob}):

\begin{eqnarray}
\label{direct} {\cal T }=\int_{\cal F} {d^2\tau \over \tau_2^{14}} {1\over
\eta^{24 }(\tau) {\bar \eta}^{24 }(\bar \tau)}\ , \\ \label{bottle}{\cal K } =
{1\over2}\int_0^\infty {d\tau_2 \over \tau_2^{14}} {1\over \eta^{24
}(2i\tau_2)}\ ,
\\ \label{annulus}{\cal A}={n^2\over2}\int_0^\infty {d\tau_2 \over
\tau_2^{14}}  {1\over
\eta^{24 }(i\tau_2/2)}\ , \\ \label{mob} {\cal M } =  {\epsilon~n
\over2}\int_0^\infty
 {d\tau_2 \over \tau_2^{14}}  {1 \over {\hat \eta}^{24 }(i\tau_2/ 2 +
1/2)}\ ,
\end{eqnarray}
where $\cal F$ is a fundamental domain of the modular group for the torus
and $\eta(\tau)$ is the Dedekind function:
\begin{equation}
\eta(\tau) = q^{1\over 24} \prod_{m=1}^{\infty} (1-q^m),~~~~~~  q=e^{2\pi i
\tau},~~~~~\tau =\tau_1 +i \tau_2\ .
\end{equation}
The `hatted' Dedekind function in ${\cal M}$ means that
the overall phase is dropped  in
$\eta(i\tau_2/ 2 + 1/2)$ ensuring that $\hat \eta(i\tau_2/ 2 + 1/2)$ is real.
A similar notation will be used for a generic
function $f$  admitting an expansion $f= q^a \sum_{i=0}^\infty a_i q^i$,
namely
$\hat f(\tau + 1/2) = e^{-i\pi a} f(\tau + 1/2)$. The world-sheet parity
operator
defining  $\cal M$ is
$\Omega = \epsilon\  (-1)^N $ where $N$ is the
open string oscillator number operator.   The
plus or minus sign $\epsilon$   in Eq.(\ref{mob}) encodes the action of
$\Omega$
on the vacuum and the one-half  shift in the argument  of the Dedekind function
in Eq.(\ref{mob}) encodes the action of the twist operator $ (-1)^N $.

The amplitudes ${\cal T}/2 \ +\ {\cal K}$ and ${\cal A} \ +\ {\cal M}$ are
respectively  the partition function of the closed and open unoriented string
sectors.    The
$q$-independent term in the expansion of the integrand of ${\cal A}+{\cal
M}$    gives the number of
massless vectors and determines the nature of the Chan-Paton group.
Using Eqs.(\ref{annulus}) and  (\ref{mob}) one finds
$n( n
-\epsilon  )/2$ massless vectors: if $\epsilon=+1$ (resp. $\epsilon =-1$), the
Chan-Paton group is
$SO(n)$ (resp.
$USp(n)$) .

To impose the tadpole condition we interpret  ${\cal K}$,  ${\cal A}$ and
${\cal M}$ as amplitudes  in the transverse  (tree) channel.  To this effect we
first define  $t=2 \tau_2$ (resp. $t=\tau_2 /2$) in
${\cal K}$ (resp. ${\cal A}$) and express the modular form $\eta^{24}(it)$ in
the integrand of Eqs. (\ref{bottle}) and (\ref{annulus}) in terms of its
$S$-transform.  The change of variable $l=  1/t$ yields
\begin{eqnarray}
\label{trans1}
  {\cal K } _{tree} &=& {2^{13}\over2}\int_0^\infty dl {1\over \eta^{24 }(il)}
 \ ,\\\label {trans1p}
\hbox{and}\qquad {\cal A } _{tree} &=&{n^2\ 2^{-13}\over2}\int_0^\infty dl
{1\over
\eta^{24 }(il)}\ .
\end{eqnarray}
The subscript $tree$ emphasises that the expressions
Eqs.(\ref{trans1}) and  (\ref{trans1p}), although identical to the
integrals  Eqs.(\ref{bottle}) and (\ref{annulus}), are now rewritten in
terms of
tree level intermediate states. Throughout the paper, any amplitude
 formulated in the `direct' channel as a one loop amplitude $\cal I$, will be
relabeled ${\cal I}_{tree}$ when expressed in terms of transverse channel tree
level intermediate states.

It is a little bit more tricky to go from the direct to   the
transverse channel for the M{\"o}bius amplitude.  One   expresses ${\hat
\eta}^{24 }(i\tau_2/ 2 + 1/2)$ in terms of its $P$-transform \cite{sagn1} which
combines the modular transformations
$S$ (i.e.
$\tau
\rightarrow -1/ \tau$) and $T$ (i.e. $\tau
\rightarrow \tau +1$):
\begin{equation}
\label{ptra} P=T^{1/2}ST^2S T^{1/2}\ .
\end{equation}
One then performs the change of variable $l=1/ (2\tau_2)$ to get
\begin{equation}
\label{trans2} {\cal M } _{tree} =  2{\epsilon~ n
\over2}\int_0^\infty dl {1\over
\hat\eta^{24 }(il + 1/2)}\ .
 \end{equation}

The tadpole condition can now be imposed by requiring the vanishing of
the  $e^{-2\pi l}$-independent term in the integrand of the  total
tree amplitude
$  {\cal K } _{tree}+  {\cal A } _{tree}+  {\cal M } _{tree}$.  One gets the
following condition
\begin{equation}
\label{not1} (2^{13}+2^{-13}n^2 - 2 \epsilon\  n)=2^{-13}(2^{13} - \epsilon\
n)^2 =0\ ,
\end{equation}
which singles out $\epsilon =+1$ and        the value $n= 2^{13}$.

  Therefore,  one recovers that the uncompactified
 open bosonic string theory obeying the tadpole condition is unoriented and
has an $SO(2^{13})$ Chan-Paton group \cite{sob}.

\subsection{Compactification of the open string ancestor on Lie algebra
lattices and truncation}

We now explain how the open string theories in 10 dimensions  may be
obtained by truncation from the compactified 26-dimensional bosonic string.
We begin by  discussing the construction of open string descendants from
closed strings  compactified on the EN lattice of a  semi-simple Lie
group ${\cal G} $ of rank $d$. We  specialise to the rank sixteen
groups
${\cal G} = E_8
\times E_8$ and
${\cal G} = E_8
\times SO(16)$ and perform the truncation.

\subsubsection{Compactification on an EN lattice}
Consider  the toroidal compactification of the  bosonic closed string theory
in 26 dimensions  on  the EN lattice  of  a  semi-simple Lie group ${\cal
G}$ of rank $d$.

The lattice partition function of the oriented closed string theory, whose
gauge symmetry is ${\cal G}\times {\cal G}  $, is
\begin{equation}
\label{part}
\sum_{\beta=1}^{\cal N}\bar\gamma_{\beta,L}(\bar
\tau)\gamma_{\beta,R}(\tau)
\ ,
\end{equation}
where the summation extends over the ${\cal N}$ conjugacy classes of the
lattice.  ${\cal N}$  is equal to the order of the centre of the universal
covering group of  ${\cal G}$. We shall hereafter label the root lattice of
${\cal G}$ by
$\beta=1$.

Unoriented closed strings are obtained by acting on the closed string
states  with the world-sheet parity projection operator  $(1+\Omega_c)/2 $
where $\Omega_c$ interchanges the quantum numbers of the left and right
sectors : $\Omega_c \vert L,R\rangle = \vert R,L\rangle$. The  Klein bottle
amplitude ${\cal K}$ is obtained from the torus  by inserting the  operator
$\Omega_c /2 $. From Eq.(\ref{part}), we see that the lattice contribution to
${\cal K}$ is the sum over the
${\cal N}$   conjugacy class partition functions
$\gamma_\beta$. Combining the lattice partition function with the partition
function of the oscillator states in the non-compact dimensions, we  get,
instead of  Eq.(\ref{bottle}),
\begin{equation}
\label{Kg} {\cal K}=
{1\over2}\int_0^\infty {d\tau_2 \over \tau_2^{14-d/2}} {1\over
\eta^{24-d }(2i\tau_2)}
\sum_{\beta=1}^{\cal N} \gamma_\beta(2i\tau_2)\ .
\end{equation}
Other choices for the lattice contribution to $\cal K$  are possible. They
differ from the one in Eq.(\ref{Kg}) by  minus signs in the sum over the
conjugacy classes in a way compatible with the fusion rules
\cite{sagn1,sagn3}.  They will not be considered here.

To compute the transverse Klein bottle
amplitude $ {\cal K}_{tree}$, we need the S-transform of
$\sum_{\beta=1}^{\cal N}\gamma_\beta(\tau)$. From  Eq.(\ref{inversion}) we
get
\begin{equation}
\label{Sweight}
\sum_{\beta=1}^{\cal N}\gamma_{\beta }(\tau)= \sqrt {\cal N} \gamma_1
(-{1\over \tau})\ .
\end{equation}
Thus, from Eq.(\ref{Kg}),
\begin{equation}
\label{tKg} {\cal K}_{tree} =  {2^{13}\over2} {\sqrt{\cal N} \over 2^{d/2}}
\int_0^\infty dl {1\over \eta^{24-d}(il)}~~
 \gamma_1(il)\ .
\end{equation}
The torus and the Klein bottle amplitudes, which only have  closed string
intermediate states,      are not altered by Chan-Paton multiplicities and we
shall
 restrict ourselves to  EN
compactifications with Klein-bottle amplitude $\cal K$ given by Eq.(\ref{Kg}).
The corresponding
${\cal K}_{tree}$ is given by Eq.(\ref{tKg}).  For such  $\cal K$,  it will
be possible to eliminate the dilaton tadpole.

We now turn to the open string sector.

The direct annulus amplitude generalising Eq.(\ref{annulus}) in absence of
Chan-Paton multiplicity   only contains the root partition function
$\gamma_1$ (see Appendix {\bf A}). It is thus given by
\begin{equation}
\label{Ag} {\cal A}= {1\over2}\int_0^\infty {d\tau_2 \over
\tau_2^{14- d /2}}  {1\over \eta^{24-d }(i\tau_2/ 2)}~~
\gamma_1 (i \tau_2 / 2) \ .
\end{equation}
The corresponding transverse amplitude, representing the exchange
of closed string modes between two boundaries, follows from the
$S$ transformation  Eq.(\ref{Sweight}),
\begin{equation}
\label{tAg} {\cal A}_{tree} =  {2^{-13 }\over 2}{2^{d/ 2}\over\sqrt{\cal
N}}\int_0^\infty dl {1\over \eta^{24-d}(il)}~~
\sum_{\beta=1}^{\cal N} \gamma_\beta(il)\ .
\end{equation}

We now introduce Chan-Paton multiplicities.   The transverse annulus
amplitude is a tree amplitude with factorisable residues. Its most general
form preserving the group symmetry is given by a sum on  all  conjugacy
classes:
\begin{equation}
\label{tAnn} {\cal A}_{tree} =  {2^{-13 }\over 2}{2^{d/ 2}\over\sqrt{\cal
N}}\int_0^\infty dl {1\over \eta^{24-d}(il)}~~
\sum_{\beta=1}^{\cal N} (a_\beta)^2\gamma_\beta(il) \ ,
\end{equation}
with arbitrary coefficients $a_\beta $. We will  see below that the
normalisation chosen in Eq.(\ref{tAnn})  reduces
to the normalisation in Eq.(\ref{tAg}) in absence of Chan-Paton multiplicities.

The direct annulus is
obtained using the modular $S$-transform. Except in the particular case where
all coefficients
$(a_\beta)^2$ are equal, the direct amplitude  contains in addition to the
root partition function
$\gamma_1$, contributions from other conjugacy classes. The latter
differ from
$\gamma_1$ by a shift
${\bf p}_\beta$ on the weight lattice of the compactified momenta.
Therefore in general, Wilson lines must be introduced to accommodate this
momentum shift in the open strings spectrum.

We make the following ansatz for $a_\beta$
\begin{equation}
\label{ndelta}
a_\beta= \sum_{\delta=1}^{\cal N}\exp (4\pi i \alpha^\prime{\bf p}_{\beta
}.{\bf p}_{\delta}) n_\delta\ ,
\end{equation}
where the $ n_\delta$ are positive integers or zero. Using
Eq.(\ref{inversion}), we get
\begin{equation}
\label{biginversion}
{1\over \sqrt{{\cal N}}}\sum_{\beta =1}^{\cal
N} (a_\beta)^2 \gamma_\beta(il)={1\over {\cal N} }\sum_{\beta, \delta,
\sigma,\tau =1}^{\cal N}\exp [4\pi i \alpha^\prime{\bf p}_{\beta }. ({\bf
p}_{\delta}+{\bf p}_{\sigma}+{\bf p}_{\tau})]  n_\sigma n_\tau
 \gamma_\delta ({i\over l})\ .
\end{equation}
The orthogonality theorem for characters of finite groups, as applied to the
centre of the universal covering of ${\cal G}$, reads
\begin{equation}
\label{ortho}
\sum_{\beta=1}^{\cal N}\exp (4\pi i \alpha^\prime{\bf p}_{\beta
}.{\bf p}_{\sigma}) \exp (4\pi i \alpha^\prime{\bf p}_{\beta
}.{\bf p}_{\tau})= {\cal N}\, \delta_{({\bf p}_{\sigma}+{\bf p}_{\tau}+{\bf
p}) }\ ,
\end{equation}
where $\sqrt{2\alpha^\prime}{\bf p} $ is an arbitrary vector of the root
lattice.   Eq.(\ref{ortho}) applied to Eq.(\ref{biginversion})  gives
\begin{equation}
\label{ouf}
{1\over \sqrt{{\cal N}}}\sum_{\beta =1}^{\cal
N} (a_\beta)^2  \gamma_\beta(il) = \sum_{\delta =1}^{\cal
N} \sum_{\sigma,\tau =1}^{\cal
N}  n_\sigma n_\tau\,
 \gamma_\delta\, ({i\over l})\quad,\quad {\bf p}_{\sigma}+{\bf p}_{\tau}+{\bf
p_\delta} +{\bf p}=0\ ,
\end{equation}
from which the direct amplitude $\cal A$ follows:
\begin{eqnarray}
\label{Aha} {\cal A}= {1\over2}\int_0^\infty {d\tau_2 \over
\tau_2^{14- d /2}}  {1\over \eta^{24-d }(i\tau_2/ 2)}~
\sum_{\delta =1}^{\cal
N} \sum_{\sigma,\tau =1}^{\cal
N}  n_\sigma n_\tau\,
 \gamma_\delta\, (i\tau_2/ 2)\ , \\
{\bf p}_{\sigma}+{\bf p}_{\tau}+{\bf
p}_{\delta} +{\bf p}=0\ .\nonumber
\end{eqnarray}
Let us illustrate this result by applying it to ${\cal
G}=SO(4m)$ (the characterisation of the conjugacy classes of $SO(4m)$ by
shift vectors was given in Section {\bf 2.2}). We get, using the constraint
in Eq.(\ref{ouf}),
\begin{eqnarray}
\label{bof}
  \sum_{\delta,\sigma,\tau =1}^{\cal
N}  n_\sigma n_\tau
 \gamma_\delta ({i\over l})= (n_o^2 +n_v^2+n_s^2+n_c^2)
 \gamma_{(o)} ({i\over l}) +
(2n_on_v+2n_s n_c)
\gamma_{(v)} ({i\over l}) \nonumber\\  +(2 n_on_s+2n_sn_c)
\gamma_{(s)} ({i\over l})    + (2 n_on_c+2n_vn_s)
\gamma_{(c)} ({i\over l})\ .
\end{eqnarray}
Eq.(\ref{bof}) exemplifies two important  features:  first, the
coefficient of the root partition function is a sum of squares; second, the
coefficients in all other conjucacy class partition functions are sums of
products $n_\sigma n_\tau ,\sigma\neq\tau$. This indicates that
$n_o,n_v,n_s,n_c$ label distinct Chan-Paton multiplicities.

Starting from the annulus amplitude $\cal A$ given in Eq.(\ref {Aha}) we  now
construct the M\"obius  amplitude. The M\"obius
amplitude in the direct channel $\cal M$  only  contains strings with identical
Chan-Paton indices at both ends.  If $\cal A$  exhibits the two features
illustrated in the example Eq.(\ref{bof}),  $\cal M$ only  contains terms
pertaining to the root lattice. Given the ansatz Eq.(\ref{ndelta}), such
features
arise if and only if ${\bf p}_{\alpha}+{\bf p}_{\beta}={\bf p}$ for any
$\alpha$ and $\beta$, as can be seen by comparing Eq.(\ref{ouf}) with
Eq.(\ref{bof}). Equivalently,
they arise if and only if the centre of the covering group of $\cal G$
only contains elements of order less or equal to two.

We shall only consider hereafter  EN compactifications on such groups.
The simple simply laced Lie groups obeying this condition are
listed in Table I and we restrict our considerations to such groups or direct
products thereof.
 \begin{center}
\begin{tabular}{||c|c|c||c|c||}
\hline
{\bf Group} &{\bf Rank ($=d$) }&{\bf Center}&$
\sqrt{\cal N}/2^{d/2}$&Rank of $b_{ab}$
\\
\hline
$SU(2)$&1& $Z_2$&$2^0$&0\\
\hline
$ SO(4m)\  m>1$&$2m$&$Z_2\times Z_2$&$2^{1-m}$&$2m-2$\\
\hline
$E_7$&7 &$Z_2$&$2^{-3}$&6\\
\hline
$E_8$&8 &$1$&$2^{-4}$&8\\
\hline
\end{tabular}

{\bf Table I}
\end{center}
We see that for all these groups, (and for their direct products),
the  factor  $\sqrt{\cal N}/2^{d/2}$ entering
Eq.(\ref{tKg}) and (\ref{tAg}) has the form $2^{-x}$ where $x$ is an integer
greater than or equal to zero. The value $x=0$ is only  realised at the
T-duality self-dual point
$[SU(2)]^d (=[SO(4)]^{d/2}$ for $d$ even).
In the last column, we relate  the factor
  $\sqrt{\cal N}/2^{d/2}$ to the rank of the antisymmetric tensor $b_{ab}$.
The latter will be used in Section {\bf 3.3} and is defined in Appendix
{\bf A}.

 To get the explicit form of $\cal M$  we must use, instead of the
standard twist
$\Omega =
\epsilon\  (-1)^N $, the group invariant  operator\footnote{The derivation of
$\Omega_{\cal G}$ in the action formalism of
Appendix {\bf A} is given in Appendix
{\bf B}.}  $\Omega_{\cal G}=
\epsilon\  (-1)^{N +\alpha^\prime{\bf p}_{op}^2}$
\cite{en},  where the eigenvalues of ${\sqrt{2\alpha^\prime}{\bf
p}_{op}}$  span the root lattice of ${\cal G}$.  The additional term
$\alpha^\prime{\bf p}_{op}^2$ in the exponent is allowed
because $\alpha^\prime{\bf p}^2 $ is integer  when $\sqrt{2\alpha^\prime}\
{\bf p}\in \Lambda_{root} $. We obtain\footnote{There are sign ambiguities
in deriving $\cal M$ from $\cal A$ given by Eq.(\ref{Aha}).  We could a priori
generalise
$\epsilon \sum_{\sigma =1}^{\cal N}  n_\sigma $ in Eq.(\ref{mobg})
to  $ \sum_{\sigma =1}^{\cal N} \epsilon_\sigma
n_\sigma $ with
$\epsilon_\sigma =\pm 1$. This arbitrariness will be lifted  by tree
channel amplitudes where Eq.(\ref{aroot}), which is true only if all
$\epsilon_\sigma$ are equal, will be needed.}
\begin{equation}
\label{mobg} {\cal M } =  {\epsilon
\over2}\int_0^\infty
{d\tau_2 \over
\tau_2^{14- d / 2}}  {1\over \hat\eta^{24-d }(i\tau_2/ 2 +
1/2)}~(\sum_{\sigma =1}^{\cal
N}  n_\sigma )~\hat\gamma_1 (i \tau_2 / 2 +
1/2) \ ,
\end{equation}
which, using  Eq.(\ref{ndelta}), can be expressed in terms of  $a_1$:
\begin{equation}
\label {aroot}
a_1=\sum_{\sigma =1}^{\cal
N}  n_\sigma \ .
\end{equation}
In Eq.(\ref{mobg}), the twist $\Omega_{\cal G} $  is encoded by shifting {\em
both} the argument of the Dedekind function and of the root lattice partition
function by
 a half unit.
Using the $P$-transform and changing variable to $l=1/ (2\tau_2)$, we rewrite
the M\"obius amplitude in terms of closed strings propagating between hole
and crosscap. For the groups $SO(8m)$ and $E_8$ we get
\begin{equation}
\label{trans2g} {\cal M } _{tree} =  2\delta_{\cal G}~{\epsilon
\over2}\int_0^\infty dl {1\over
\hat\eta^{24-d }(il + 1/2)} (\sum_{\sigma =1}^{\cal
N}  n_\sigma )~\hat\gamma_1 (i l+
1/2) \ ,
 \end{equation}
where $\delta_{\cal G}$ is a sign depending on the group $\cal G$ considered;
$\delta_{\cal G} =
+1$ for
$E_8$ and
$\delta_{\cal G} =
(-1)^m$ for
$SO(8m)$. For
$SU(2)$, $SO(8m +4)$ and $E_7$,  the situation is different.  The transverse
amplitude ${\cal M}_{tree}$ derived from Eq.(\ref{mobg}) is not proportional
to
$\hat\gamma_1$.

Two conditions have to be met for the consistency of the theory. First,
${\cal A}+{\cal M}$  must describe the partition function of unoriented open
strings with products of Chan-Paton groups $SO(n)$ or $USp(n)$ and
second,  each term in the
$e^{-2\pi l}$ power series expansion    of the integrand of the total  tree
channel amplitude ${\cal K}_{tree}+{\cal A}_{tree}+{\cal M}_{tree}$ must be a
perfect square.

With the ansatz Eq.(\ref{ndelta}), the first condition is always  satisfied for
compactification on EN lattices of the groups of Table I. The Chan-Paton
groups are determined by the massless vector contributions to  ${\cal
A}+{\cal M}$, which appear as $q$-independent terms ($q= e^{-\pi\tau_2/2}$) in
the expansion of the integrand of  Eqs.(\ref{Aha}) and
(\ref{mobg}).  These $q$-independent terms  are of two types:  terms
arising from level zero in $\eta$ and
$\hat\eta$ and level one in $\gamma_1$ and $\hat\gamma_1$ are scalars
(previously referred to as
$\cal G$-scalars), while those arising from
the  level one in $\eta$ and
$\hat\eta$ (which has degeneracy $(24-d)$) and  level zero in
$\gamma_1$ and $\hat\gamma_1$ are  vectors.  The number of such
massless vectors is
$\sum_{\sigma=1}^{\cal N} n_\sigma(n_\sigma -\epsilon)/2 $. We thus get
$\cal N$ direct products of orthogonal (resp. symplectic) Chan-Paton groups
for $\epsilon =+1$ (resp. $
\epsilon =-1$) whose total multiplicity is $\sum_{\sigma=1}^{\cal N}
n_\sigma$, provided
 the second condition is verified. This is the case for $E_8$ and $SO(8m)$ and
we further analyse the unoriented bosonic theories compactified on these
groups\footnote{For the other groups listed in Table I, one may choose
$\cal K$ different from  Eq.(\ref{Kg}) in  order to  fullfill the perfect
square
condition.}.

These  consistent theories  also have  two types of massless
particles in the closed string channel: particles created by oscillators in
non-compact dimensions and $\cal G$-scalars.  Massless
exchange generates divergences  in the
$e^{-2\pi l}$-independent terms in the expansion of the integrand of
${\cal K}_{tree}+{\cal A}_{tree}+{\cal M}_{tree}$.  There are
massless exchanges  in the terms proportional to $\gamma_1$ (and $\hat
\gamma_1$) in Eqs.(\ref{tKg}), (\ref{tAnn}) and (\ref{trans2g}).
The dilaton and the graviton are exchanged  at level zero in
$\gamma_1$ and level one in $\eta$ and massless  $\cal G$-scalars are
exchanged  at level one in
$\gamma_1$ and level zero in $\eta$. These divergences are all eliminated
by imposing the dilaton tadpole condition:
\begin{equation}
{2^{13}\over2} {\sqrt{\cal N} \over 2^{d/2}} +{2^{-13 }\over 2}{2^{d/
2}\over\sqrt{\cal
N}} a_1^2 -\delta_{\cal G} \epsilon\  a_1 = {2^{-13 }\over 2}{2^{d/
2}\over\sqrt{\cal N}}(a_1 - \delta_{\cal G}\epsilon \  2^{13}{\sqrt{\cal N}
\over
2^{d/2}})^2=0\ .
\end{equation}
Choosing $\epsilon$  such that $\delta_{\cal G}\epsilon = +1$,
we obtain from  (\ref{aroot}) the total Chan-Paton multiplicity :
\begin{equation}
\label{dtad}
\sum_{\sigma =1}^{\cal
N}  n_\sigma =  2^{13}{\sqrt{\cal N} \over 2^{d/2}}\ .
\end{equation}
In certain cases, there may  be additional zero mass $\cal G$-scalars
exchanges in ${\cal A}_{tree}$ in terms proportional to $\gamma_\sigma ,
\sigma\neq 1$. Imposing the tadpole condition for these massless $\cal
G$-scalars may lead to further restriction on the Chan-Paton groups, as will
be seen in the next section.

The  cases of interest for our analysis of truncation are $E_8\times E_8$ and
$E_8\times SO(16)$.  As $\delta_{ \cal
G}=1$ for both groups, we must take $\epsilon=1$ to eliminate the dilaton
tadpole. The Chan-Paton groups  are
\begin{eqnarray}
\label{sup} E_8\times E_8 &\Rightarrow& SO(n)\ ,\\
\label{nsup}
E_8\times SO(16)&\Rightarrow &SO(n_0)\times  SO(n_v)\times
SO(n_s)\times  SO(n_c)\ ,
\end{eqnarray}
where $n, n_{0}, n_{v}, n_{s}, n_{c}$ are arbitrary positive integers or zero.

We now construct in detail the tadpole-free theories for EN
compactification on these groups and perform the
truncation.

\subsubsection{The  $E_8 \times E_8$ compactification and Type I
superstring}

We consider here the  compactification  on the
$E_8 \times E_8$ lattice. In this case  there is  only one conjugacy class
( ${\cal N}=1$). We write
\begin{equation}
\gamma_1 =\gamma_{{(o)}_{E_8}}
\gamma_{{(o)}_{E_8}}=(\gamma_{{(o)}_{16}} +
\gamma_{{(s)}_{16}}) (\gamma_{{(o)}_{16}} +
\gamma_{{(s)}_{16}})\ .
\end{equation}

The tree amplitudes  given by Eqs.(\ref{tKg}),
(\ref{tAnn})    and (\ref{trans2g}),  with   $\delta_{ \cal
G}=\delta_{ E_8\times E_8}=1$,  read
\begin{eqnarray}
\label{ksup} {\cal K } _{tree}& =& {2^5 \over2}\int_0^\infty dl~~ {1\over
\eta^{8 }(il)}
\gamma_{{(o)}_{E_8}}(il) \gamma_{{(o)}_{E_8}}(il)\ ,
\\ \label{asup}{\cal A } _{tree}& =&{n^2\ 2^{-5}\over2}\int_0^\infty dl~~
{1\over \eta^8 (il)}
\gamma_{{(o)}_{E_8}}(il) \gamma_{{(o)}_{E_8}}(il)\ ,\\ {\cal M } _{tree}& =&
\epsilon n \int_0^\infty dl~~ {1 \over {\hat \eta}^8 (il+{1 \over 2})} {\hat
\gamma}_{{(o)}_{E_8}}(il+1/ 2)  {\hat \gamma}_{{(o)}_{E_8}}(il+ 1
/ 2)\ .
\label{msup}
\end{eqnarray}
They can easily be  expressed as  direct channel amplitudes ${\cal K}$, ${\cal
A}$ and ${\cal M}$.

We       impose the vanishing  of the dilaton tadpole.  All the transverse
amplitudes are  proportional to the root lattice of  $E_8\times E_8$.
Therefore, as explained in the preceding section,
{\em all} divergences  are
 eliminated by the
 single constraint
 Eq.(\ref{dtad}).  This yields
\begin{equation}
\label{not2}  n=2^5 \ ,
\end{equation}
and $\epsilon = +1$.
The Chan-Paton group  is given by  Eq.(\ref{sup}). It is
$SO(32)$. Let us stress again that this reduction of
Chan-Paton group from $SO(2^{13})$ by compactification is crucially related
to the structure of the Lie algebra lattice, and in particular to  the
number of
conjugacy classes, as shown in
 Eq.(\ref{dtad}).

We are now in position to derive the truncated theory from the
tadpole-free
 bosonic open string theory  compactified
on the EN lattice of $E_8 \times E_8$.

Eq.(\ref{truncations})
gives
\begin{center}
$\gamma_{(o)_{E8}}\gamma_{(o)_{E8}}
=  \gamma_{(o)_{E8}}(\gamma_{(o)_{16}}+
\gamma_{(s)_{16}})$

Trunc $\downarrow$ Flip

$\gamma_{(v)_{8}}-\gamma_{(s)_{8}}$
\end {center}
and the transverse amplitudes Eqs.(\ref{ksup}), (\ref{asup}) and (\ref{msup})
become after truncation
\begin{eqnarray}
\label{Ktt} {\cal K }^{t} _{tree}& =& {\cal A}^{t}_{tree} =  {2^5 \over
2}\int_0^\infty dl~~ {1\over \eta^8(il) }
(\gamma_{(v)_{8}}-\gamma_{(s)_{8}})(il) \ ,
 \\ \label{Mtt}  {\cal M }^{t} _{tree}& =& - 2^5 \int_0^\infty dl~~ {1
\over {\hat
\eta}^8 (il +1 / 2) } (\hat \gamma_{(v)_{8}}  - \hat \gamma_{(s)_{8}})
 (il +1 / 2) \ .
\end{eqnarray}
The flip in sign  in Eq.(\ref{Mtt}) as compared to
Eq.(\ref{msup})  with $\epsilon=+1$ is solely due to the definition of hatted
functions. Truncating the direct amplitudes ${\cal A}$ and ${\cal M}$ we get
\begin{eqnarray}
 \label{At}
{\cal A}^{t}&=& {2^{10}\over2}\int_0^\infty {d\tau_2 \over \tau_2^6}
{1\over \eta^8 (i\tau_2/ 2)}
( \gamma_{(v)_{8}} - \gamma_{(s)_{8}})(i\tau_2/ 2)\ ,\\ \label{Mt} {\cal M}^t&
=& - {2^5
\over 2} \int_0^\infty  {d\tau_2 \over \tau_2^6}
{1\over {\hat \eta}^8 (i\tau_2/ 2+1 /2)} (\hat \gamma_{(v)_{8}} -
\hat\gamma_{(s)_{8}})(i\tau_2/ 2+1 /2)\ .
 \end{eqnarray}
The amplitudes in  Eqs.(\ref{At}) and (\ref{Mt}) are equal to those in
 Eqs.(\ref{Ktt}) and (\ref{Mtt}) expressed in the transverse channel, a
 consequence of the fact that the truncation-flip commutes with $S$ and $T$
 transformations.

Massless modes in both open and closed string channels, created by bosonic
oscillators in non compact dimensions, become massive after truncation.  A
subset of massless $\cal G$-scalars are transmuted to massless modes.
 In the
closed string  channel, these are  massless spinors,  $NS$-$NS$ and
$R$-$R$ fields. The $NS$-$NS$ and $R$-$R$   tadpoles  are
 eliminated by the  condition Eq.(\ref{not2}) inherited from the bosonic
string.
This is easily checked from Eqs.(\ref{Ktt}) and
(\ref{Mtt}) describing  the truncated tree amplitude ${\cal K }^{t}
_{tree}+{\cal A }^{t} _{tree}+{\cal M }^{t} _{tree}$. Note that the flip in the
$\epsilon$ sign in Eq.(\ref{Mtt})  is in accordance with
$\delta_{\cal G}\epsilon = +1$, as $\delta_{SO(8)} =-1$. This flip in
sign translates the fact that the zero mass states which contribute to
the truncated tree amplitudes Eqs.(\ref{Ktt}) and (\ref{Mtt}) stem
from  the lattice partition functions at level zero, while the  ${\cal
G}$-scalars in the transverse amplitudes of the untruncated bosonic theory
 given in Eqs.(\ref{ksup}),
(\ref{asup}) and (\ref{msup}) arose at level one.

The Chan-Paton group  $SO(32)$ is preserved under truncation. This can be
checked  by counting, in the open channel,  the number of massless vectors
 arising by truncation from ${\cal G}$-scalars. This number  (which is the
same as the number of massless spinors) is
$2^5(2^5 - 1)/2$, as is  immediately apparent from the partition
function  ${\cal A}^t + {\cal M}^t$  given by Eqs.(\ref{At}) and
(\ref{Mt}). The
latter is then the partition function of the open string sector of Type I.
Note
that the $2^5(2^5 - 1)/2$  vectors  of the untruncated theory stem from the
Dedekind function at level one in ${\cal A} + {\cal M}$ while the vectors of
the truncated theory stem from level zero in ${\cal A}^t + {\cal M}^t$, a
manifestation of the change in the Lorentz group in truncation which allows
the transmutation of
$\cal G$-scalars  to tensors. Similar remarks to those
 concerning the flip of  $\epsilon$ sign in the closed string channel can be
made in the open channel.

We see that the  truncation of the unoriented tadpole-free bosonic open
string theory, compactified on the
$E_8
\times E_8$  lattice,  results in  the $SO(32)$ anomaly free Type I theory.

\subsubsection{The $E_8 \times SO(16)$ compactification and Type 0
strings}

We now consider  the 26-dimensional  unoriented bosonic open string
theory compactified on the $E_8 \times SO(16)$ lattice. In this case there
are {\it four} conjugacy classes,
${\cal N}=4$:
$\gamma_{{(o)}_{E_8}} \gamma_{(o)_{16}},  \gamma_{{(o)}_{E_8}}
\gamma_{(v)_{16}},~\gamma_{{(o)}_{E_8}} \gamma_{(s)_{16}},~
\gamma_{{(o)}_{E_8}} \gamma_{(c)_{16}}$ and we have
\begin{equation}
\gamma_1=\gamma_{{(o)}_{E_8}}\gamma_{{(o)}_{16}} \  .
\end{equation}
The tree amplitudes  given by Eqs.(\ref{tKg}),
(\ref{tAnn}),    and (\ref{trans2g}) , with  $\delta_{
E_8\times SO(16)}=1$, read
\begin{eqnarray}
\label{ESt} {\cal K } _{tree} &=& {2^6 \over2}\int_0^\infty dl {1\over
\eta^8 }
\gamma_{{(o)}_{16}} \gamma_{{(o)}_{E8}}\ ,
\\ \label{ASt}{\cal A } _{tree} &=&{2^{-6}\over2}\int_0^\infty dl {1\over
\eta^8} (a^2_1 \gamma_{{(o)}_{16}} + a^2_2 \gamma_{{(v)}_{16}}  + a^2_3
\gamma_{{(s)}_{16}}    + a^2_4 \gamma_{{(c)}_{16}})
\gamma_{{(o)}_{E_8}},\\
\label{MSt} {\cal M } _{tree} &=& \epsilon a_1
\int_0^\infty dl {1
\over {\hat \eta}^8}  {\hat \gamma}_{{(o)}_{16}}{\hat
\gamma}_{{(o)}_{E_8}}\ ,
\end{eqnarray}
and
\begin{eqnarray}
\label{chanv} a_1&=&n_o+n_v+n_s+n_c\ , ~~~~~ a_2=n_o+n_v-n_s-n_c\ ,
\nonumber \\ a_3&=&n_o-n_v+n_s-n_c \ ,~~~~~ a_4=n_o-n_v-n_s+n_c\ .
\end{eqnarray}

We  enforce the tadpole conditions. In this case there
are four tadpole conditions because  there are four different types of
massless modes giving rise  to divergent contributions in the tree
amplitudes: in addition to the graviton and dilaton encoded in the Dedekind
function at level one, there are  three  types of $\cal G$-scalars depicted
respectively  by
$\gamma_{(o)_{16}}$ at level 1 and
$\gamma_{(s)_{16}}$,
$\gamma_{(c)_{16}}$ at level zero.
The tadpole conditions which eliminate the divergences at level one in
the Dedekind function and in
$\gamma_{(o)_{16}}$ are both given by Eq.(\ref{dtad}):
\begin{equation}
\label{tadp1}
 n_o+n_v+n_s+n_c = a_1= 64 \ ,
\end{equation}
with $\epsilon =1$.  The Chan-Paton group  is given by Eq.(\ref{nsup}). It is
  $SO(n_o) \times SO(n_v) \times SO(n_s) \times
SO(n_c)$, with
$n_o+n_v+n_s+n_c=64 $.
The pattern of symmetry breaking of $SO(64)$ is further determined by
imposing the two remaining tadpole conditions corresponding to the
$\gamma_{(s)_{16}}$ and $\gamma_{(c)_{16}}$ divergences.. They give:
\begin{eqnarray}
\label{break} a_3=0 &,&a_4=0 \nonumber  \\ &\Updownarrow& \\
 n_o=n_v &,& n_s=n_c \nonumber
\end{eqnarray}
The Chan-Paton group of the tadpole-free unoriented bosonic open string
theory compactified on the
$E_8 \times SO(16)$ lattice is thus:
\begin{equation}
\label{grob} SO(n) \times SO(n) \times SO(32-n) \times SO(32-n)\ .
\end{equation}
We now derive the truncated theory.

Using Eq.(\ref{truncations}),  the truncation-flip
\begin{eqnarray}
\gamma_{(o)_{16}} \rightarrow+ \gamma_{(v)_{8}}\ , &\quad&
\gamma_{(v)_{16}} \rightarrow +\gamma_{(o)_{8}} \ , \nonumber \\
\gamma_{(s)_{16}} \rightarrow -\gamma_{(s)_{8}}\ , &\quad&
\gamma_{(c)_{16}} \rightarrow - \gamma_{(c)_{8}}\ , \nonumber
\end{eqnarray}
 applied to the amplitudes Eqs.(\ref{ESt}), (\ref{ASt})
and (\ref{MSt}) gives,
\begin{eqnarray}
\label{EStt} {\cal K }^t_{tree} &=& {2^6 \over2}\int_0^\infty dl {1\over
\eta^8 }
\gamma_{{(v)}_8} \ ,
\\ \label{AStt}{\cal A }^t_{tree} &=&{2^{-6}\over2}\int_0^\infty dl {1\over
\eta^8} [(64)^2 \gamma_{{(v)}_{8}} +
 16 (n-16)^2 \gamma_{{(o)}_{8}}] \ ,  \\ \label{KStt}{\cal M }^t_{tree} &=& -64
\int_0^\infty dl {1 \over {\hat \eta}^8}  {\hat \gamma}_{{(v)}_{8}} \ .
\end{eqnarray}
The transverse amplitudes propagate  massless $NS$-$NS$ particles at
level zero in $ \gamma_{{(v)}_{8}}$, inherited from  $\cal G$-scalars
 propagating in $\gamma_{(o)_{16}}$. The potential
$NS$-$NS$ tadpole has been eliminated  by the condition Eq.(\ref{tadp1}). The
truncated direct amplitudes
${\cal A}^t$ and
${\cal M}^t$ follow from Eq.(\ref{AStt}) and (\ref{KStt}) using $S$ and $P$
transformations or equivalently by truncating the direct amplitudes.  These
now contain  all four conjugacy classes of $SO(8)$ but $\cal M$ only contains
$\hat\gamma_{(v)_8}$. This indicates that open strings with identical
Chan-Paton factors at both ends  only  have the bosonic $NS_+$ spectrum with
the tachyon projected out,  while  open strings  with distinct Chan-Paton
factors at their ends have tachyons or fermionic states of either chirality.

As in the preceding section, the massless states in both open and closed
channels again arise from a subset of $\cal G$-scalars  and the truncation
exhibits  the concomitant flip of  $\epsilon$ sign between  ${\cal M
}_{tree} $ in
Eq.(\ref {MSt}) and  ${\cal M }^t_{tree}$ in Eq.(\ref {KStt}).  The Chan-Paton
group is transferred from the untruncated bosonic theory to the truncated
fermionic theory and  we obtain the spectrum of the
$[SO(32-n) \times SO(n)]^2$ Type O  theories discussed in
ref.\cite{sagn1,bega}.

We see that the  truncation of the unoriented tadpole-free bosonic open
string theory, compactified on the
$E_8 \times
SO(16)$  lattice,  results in  the tadpole-free 10-dimensional Type O theory
with gauge group  $[SO(32-n) \times SO(n)]^2$.  Note that the Chan-Paton
group has a higher rank  than in Type I because $E_8 \times  SO(16)$ has more
conjugacy classes than $E_8\times E_8$. We emphasise that this Chan-Paton
gauge structure and the symmetry breaking pattern arise entirely
from bosonic considerations.

\subsection{Brane fusion}

In this section we discuss the  interpretation of the
compactified $SO(2^{13})$ unoriented  bosonic theory in terms of D-branes
and orientifolds.

The uncompactified  $SO(2^{13})$  bosonic string  has a clear
geometrical interpretation \cite{pol}.  The ends of the open strings
live on  D25-branes and the tension of a D25-brane can be derived from
${\cal A}_{tree}$  (see Eq.(\ref{trans1p}) with $n=1$) by comparing with the
field theory calculation \cite{pol}.

The tension $T^{bosonic}_{D25}$ is given by\footnote{We always give  the
Dp-brane tensions  computed in the oriented closed theory.}
\begin{equation}
\label{brane} T^{bosonic}_{D25} = {\sqrt\pi\over2^4
  \kappa_{26}}(2\pi\alpha^\prime{}^{1/2})^{-14}\ ,
\end{equation}
 where $\kappa_{26}^2= 8\pi G_{26}$ and $G_{26}$ is the Newtonian
constant in 26 dimensions.

The action of the world-sheet  parity operator on the 26-dimensional
closed bosonic string introduces  an orientifold 25-plane $O25$. The
tension of the $O25$ can be derived from ${\cal K}_{tree}$ (and ${\cal
M}_{tree}$ for the sign of this tension)  again by  comparing with the field
theory calculation. The result  is \cite{pol}:
\begin{equation}
\label{orientifold}  T_{O25}^{bosonic} = - 2^{12} T^{bosonic}_{D25}\ .
\end{equation}
Therefore the tadpole condition fixing the $SO(2^{13})$ gauge group  means
in this context that one has to introduce $n=2^{13}$ D25-branes (
$2^{12}$+ their images) to cancel the negative tension of  the $O25$
\cite{pol}.

We now derive the tension of the wrapped orientifold and D25-branes for
compactifications on EN lattices with Lie group ${\cal G}$ of rank $d$,
where  ${\cal G}$ is a direct product of the groups listed in table I.
To this effect we describe the lattice in terms of  constant background
metric $g_{ab}$ and Neveu-Schwarz antisymmetric tensor $b_{ab}$. For EN
lattices, these are quantised and their explicit form  is given in appendix
{\bf A}.

The  squared wrapped orientifold tension  is obtained from the coupling to
gravity in the transverse Klein bottle amplitude ${\cal
K}_{tree}$. For the groups $E_8$ and $SO(8m)$   and their direct
products, ${\cal K}_{tree}$
contains dilaton and graviton exchange terms at level zero in
$\gamma_1$, as exhibited in Eq.(\ref {tKg}). We recall that ${\cal K}_{tree}$
is obtained from the direct  amplitude ${\cal K}$ by the S modular
transformation, and that
${\cal K}$ follows from the torus amplitude by inserting the world-sheet
parity operator $\Omega_c$ which interchanges left and right sectors.
This projects out the $b_{ab}$-field as can be checked by imposing ${\bf p_L}=
{\bf p_R}$ in Eq.(\ref{close}).  For the above groups, the tension of the
wrapped orientifold $T_{O25~ wr}^{bosonic}$ after  compactification  is then
given, up to a sign,  by:
\begin{equation}
\label{worien} T_{O25~ wr}^{bosonic}=T_{O25}^{bosonic} [(2 \pi)^d \sqrt{g}]\ ,
\end{equation}
where $g$ is the determinant of the metric. This can be verified by comparing
the zero mass propagation in Eq.(\ref {tKg}) with its field-theoric
counterpart.

On the other hand the world-volume action of a  D25-brane is given  by the
Born-Infeld action \cite{tens1,tens2,cal}. Accordingly,  after
compactification,  the tension $T_{D25~ wr}^{bosonic}$ of a wrapped
D25-brane is given  by \cite{hkms}:
\begin{equation}
\label{wdbrane}  T_{D25~ wr}^{bosonic}=T_{D25}^{bosonic} [(2 \pi)^d
\sqrt{e}] = n_f T_{D25}^{bosonic} [(2 \pi)^d
\sqrt{g}]\ ,
\end{equation} where $e$ is the determinant of $e_{ab} \equiv
g_{ab}+b_{ab}$, and
\begin{equation}
\label{reduc}
n_f = \sqrt{ e /g}\ .
\end{equation}
The choice $\delta_{\cal G} \epsilon = +1$ in
the transverse M\"obius amplitude Eq.(\ref{trans2g}), which allows
the cancellation of the dilaton tadpole divergence, means that the
tension   of the wrapped orientifold Eq.(\ref{worien})
 is negative.  This tension can then be compensated   by introducing
$n_w$ wrapped D25 branes (including images) with:
\begin{equation}
\label{crit} n_w =  2^{13} \sqrt{g \over e} =
{2^{13}\over n_f} \ .
\end{equation}
The reduction  factor $n_f$  defined in Eq.(\ref{reduc}) is computed in
Appendix {\bf C}. One gets
\begin{equation}
\label{fuse}  n_f ={2^{d/ 2} \over \sqrt
{\cal N}}\ .
\end{equation}
Inserting this value in Eq.(\ref{crit}), we recover the dilaton tadpole
condition
Eq.(\ref{dtad}), with the total Chan-Paton
multiplicity equal to the number $n_w$ of wrapped D-branes
required for cancelling the wrapped orientifold tension.  In other words, when
the tadpole condition is imposed,  the  reduction factor $n_f$ coincides with
the ratio of Chan-Paton multiplicities before and after compactification.

Recall that for the  Lie groups ${\cal G}$
listed in table I and direct products thereof, the factor $2^{d/ 2}/ \sqrt
{\cal N}=n_f $  has the form
$2^q$, with $q$ a positive integer or zero.  For compactifications on EN
lattices, the integer $q$ is related to the rank
$r$  of the
$b_{ab}$ matrix listed in Table I. We have indeed
\begin{equation}
\label{rank}
{2^{d/2}\over\sqrt{\cal N}}= 2^{r/2} \ ,
\end{equation}
as can be verified by direct inspection from the explicit form of $b_{ab}$
given in Appendix {\bf A}.

One may interpret the reduction of the total Chan-Paton
multiplicity as a consequence of  `brane fusion'.  Eq.(\ref{wdbrane}) shows
that {\em one} D25-brane of the 26-dimensional bosonic string theory,
compactified on EN lattices with
${\cal G}$ in Table I,  has the same tension as the integer number
$n_f=\sqrt{e/g}$ of D25-branes when the  theory is compactified on a
generic Cartesian torus (for which
$b_{ab}=0$) with  equal volume. This fact does not depend on the
orientability of the string nor on the tadpole condition. In presence of the
enhanced symmetry, the  r\^ole of the quantised $b_{ab}$-field is  to  fuse
$n_f$ component branes into {\em one} .
If the  tadpole condition is
imposed for groups with factors $E_8$ and $SO(8m)$, one gets the
concomitant reduction in the Chan-Paton multiplicity from the  unoriented
26-dimensional bosonic string   theory with Chan-Paton group
$SO(2^{13})$. Such reduction, and its expression in terms of   the rank of the
$b_{ab}$ matrix,   is in agreement with reference
\cite{sagn2}.

We now specialise to $d=16$ and apply these results to $E_8 \times E_8$
and $E_8 \times SO(16)$.

For $E_8 \times E_8$ we have ${\cal N}=1$ and  Eq.(\ref{fuse}) gives
$n_f = 2^8$. Consequently
$2^8$ D25-branes $\rightarrow$ one D25-brane  and  we are left with $2^5$
 wrapped D25-branes.  This is consistent
with  the fact that the Chan-Paton group of the tadpole-free
$E_8 \times E_8$ compactification is  $SO(32)$.

For $E_8 \times SO(16)$ we have  ${\cal N}=4$. Thus Eq.(\ref{fuse}) gives
$n_f = 2^7$. Consequently
$2^7$ D25-branes $\rightarrow$ one D25-brane and  we are left with
$2^6$ wrapped  D25-branes. This is consistent with  the fact that
the Chan-Paton group of the tadpole-free
$E_8 \times SO(16)$ compactification is  $[SO(32-n) \times SO(n)]^2$.

\subsection{Energy conservation in the truncation}

We now show that truncation conserves energy. Namely  the
orientifold and  D-brane tensions after  truncation are equal to the
ten-dimensional  orientifold and  D-brane tensions of the
compactified  bosonic ancestor.

We first discuss the orientifold tensions. Before truncation, the square of
the 10-dimensional tension  of the bosonic wrapped orientifold is
proportional to the
$q^0$-term ($q=e^{-2 \pi
l}$) in the integrand of ${\cal K}_{tree}$ due to the
exchange of massless space-time modes.  Using Eqs. (\ref{ksup}) and
(\ref{ESt}), we have  for both
$E_8 \times E_8$ and $E_8 \times SO(16)$:
\begin{eqnarray}
\label{tow} {\cal K }_{tree}\vert_{q^0} &=& 2^4 ~ {\sqrt {\cal
N}}~~~~\eta^{-8} (il)\vert_{{\rm level}~1}~~ \gamma_1(il)
\vert_{{\rm level}~0} \nonumber\\ &=& 2^4 ~ {\sqrt {\cal N}} ~~~~ q^{-8/
24} (8q)~~~~~ q^{-16/ 24}~~=~ 2^4~ {\sqrt {\cal N}}~8\ .
\end{eqnarray}
 After truncation the square of     orientifold tension is
proportional  to the $q^0$-term in the integrand of    ${\cal
K}^t_{tree}$ coming from
$\gamma_{(v)}$.
 Using Eqs. (\ref{Ktt}) and  (\ref{EStt}), we have
 for both
$E_8 \times E_8$ and $E_8 \times SO(16)$
\begin{eqnarray}
\label{tot} {\cal K }_{tree}^{t}\vert_{q^0} &= &  2^4 ~ {\sqrt
{\cal N}}~~~~\eta^{-8} (il)\vert_{{\rm level}~0}~~
\gamma_{(v)_{8}}(il)\vert_{ {\rm level}~0}
\nonumber \\ &=&  2^4 ~ {\sqrt {\cal N}}~~~~q^{-8/ 24}~~~~~~~ [q^{-4/
24} (8 q^{+1/ 2})] =  2^4 ~ {\sqrt {\cal N}}~8\ .
\end{eqnarray}
The  amplitudes Eq.(\ref{tow}) and Eq.(\ref{tot}) are independent of coupling
constants. Their equality implies therefore that the
orientifold tension $T_{09}^{ferm}$ generated by truncation  is equal
to  the tension of the wrapped bosonic orientifold $T_{O25~
wr}^{bosonic}$ when expressed in terms
of the same 10-dimensional Newton constant $\kappa_{10}$:
\begin{equation}
\label{kappadix}
\kappa_{10}=\kappa_{26} ( (2\pi)^{16} \sqrt{g})^{-1/2} =
\kappa_{26}(2\pi \alpha^\prime{}^{1/2})^{-8}\,  2^4 {\cal N}^{-1/4}\ .
\end{equation}
The conservation of the   orientifold tension in the truncation process
implies in turns that the 10-dimensional tension of the D-branes is
conserved. Indeed the truncated theories being tadpole-free, the total
tension  of the $n_w$ D9-branes generated by truncation from the
compactified theory with $n_w$ wrapped D25-branes must cancel  the
orientifold tension. Using Eqs.(\ref{kappadix}), (\ref{brane}),
(\ref{wdbrane}) and (\ref{eg}) the tension $T_{D9}^{ferm}$ of   D9-branes
in the truncated theory is thus predicted to be
\begin{equation} T_{D9}^{ferm}=T_{D25~wr}^{bosonic}={\sqrt\pi\over 2^4
 \kappa_{26}}(2\pi\alpha^\prime{}^{1/2})^2= {\sqrt\pi\over
 \kappa_{10}}(2\pi\alpha^\prime{}^{1/2})^{-6} {\cal N}^{-1/4}\ .
\label{tenf}
\end{equation}
For $E_8 \times E_8$, ${\cal N} =1$ and Eq.({\ref{tenf}}) gives
\begin{equation}
\label{teniib} T_{D9}^{ferm}= {\sqrt\pi\over
 \kappa_{10}}(2\pi\alpha^\prime{}^{1/2})^{-6}\ ,
\end{equation} which is indeed the tension of the Type IIB D9-brane (see
\cite{pol}).

For $E_8 \times SO(16)$, ${\cal N} =4$ and Eq.({\ref{tenf}}) gives
\begin{equation}
\label{tenob} T_{D9}^{ferm}= {\sqrt\pi\over
\sqrt{2} \kappa_{10}}(2\pi\alpha^\prime{}^{1/2})^{-6}\ ,
\end{equation} which is indeed the tension of the Type OB D9-branes (see
for instance
\cite{tseyz}).

\section{Discussions}

Although truncation  is a universal procedure to extract
all  ten-dimensional closed fermionic string
 theories out  of the closed
bosonic string, one might have the feeling that such a procedure is
essentially kinematical in character.  One  may  always bosonise the
world-sheet fermions to obtain, in the light-cone gauge,  theories formally
identical to those resulting from truncation\footnote{See
\cite{revlsw} and references therein.}.  The added information provided by
truncation  is the encoding of  all these theories   in
the 26-dimensional bosonic string theory compactified,  in the left and/or
right sector, on the Lie algebra lattice of $E_8\times SO(16)$.  In particular,
truncation selects the supersymmetric  strings  by extracting  the only
sublattice of $E_8\times SO(16)$  (namely  $E_8\times E_8$)  which   yields a
distinct compactification of the bosonic string consistent with  modular
invariance. This emergence of all fermionic strings  from subspaces
of the bosonic string Hilbert space in a systematic way  is certainly
impressive,  but it is unclear that  truncation can predict specific properties
of the fermionic strings.

As a first step to convince oneself that truncation hides some dynamics, one
would like to find such predictions.

This was the motivation which led us to generalise truncation to  open
strings. There, crucial ingredients of the fermionic strings, such as the
cancellation of anomalies in Type I theory by the Chan-Paton group
$SO(32)$, a priori seemed to bear no relation with the bosonic theory.
Remarkably however,  through the brane fusion mechanism,  the imposition
of a tadpole condition in the bosonic string  explains this anomaly
cancellation.  It also yields the Chan-Paton groups which eliminate all
tadpoles in Type O theories. In addition, the  tensions of the fermionic
space-filling D-branes are computed using   properties of the bosonic
string only. More important even is the fact that these results follow from
specific properties of the $E_8\times SO(16)$ lattice, singling out this
lattice out of  all possible toroidal compactifications.

We think that this host of results points towards a dynamical origin of
truncation. But such dynamics cannot be handled in the context of the
perturbative approach given here. To settle the issue, a manageable
non-perturbative approach to the bosonic string is mandatory. An attempt
towards formulating such an approach in a classical limit has been recently
proposed \cite{suss} and the present analysis suggests that such efforts
should be further pursued.

If truncation is indeed a dynamical phenomenon,  the central r\^ole played
by the 26-dimensional bosonic string theory as a parent of all the fermionic
strings would have important consequences for quantum gravity. Whatever
may ultimately happen to string theory itself,  it would probably mean that
the fundamental constituents of quantum gravity do not contain fermionic
degrees of freedom.
\newpage

\section*{Acknowledgments}

 Fran\c cois  Englert is grateful to Peter West for a fruitful conversation.
Laurent Houart thanks Augusto Sagnotti for illuminating discussions.  We
thank Auttakit Chattaraputi and Alexander Sevrin for constructive
comments.  F.E. acknowledges EPSRC for a  Visiting Fellowship
GR/N22793/01.  A.T. acknowledges the Leverhulme Trust for a fellowship.
This work is supported by the European Commission RTN program
HPRN-CT-2000-00131 in which F.E. is associated to the University of
Brussels (ULB) and L.H. is associated to the University of Milano Bicocca
(Milano 2).
\newpage
\appendix
\setcounter{equation}{0}
\renewcommand\theequation{\thesection.\arabic{equation}}

\section{Toroidal compactification on EN lattices}
 Toroidal compactification of the bosonic string is described by
constant metric and  antisymmetric tensors. The action is \cite{narain}
\begin{eqnarray} S={-1\over 4 \pi \alpha^\prime}\int d\sigma d\tau
\left[\{g_{ab}\partial_\alpha X^a\partial^\alpha X^b + b_{ab}
\epsilon^{\alpha \beta} \partial_\alpha X^a\partial_\beta X^b\}\right.
\nonumber\\
\left. +\eta_{\mu\nu}\partial_\alpha X^\mu\partial^\alpha X^\nu
\right]\ ,
\label{action}
\end{eqnarray} with $g_{ab},b_{ab} $ a constant metric and antisymmetric
tensor in compact directions ($a,b=1,...d$),\\
$\eta_{\mu\nu},\eta_{\alpha\beta} =
(-1;+1,...),~~~~~~\mu, \nu =1,...,s+2, ~~~\alpha, \beta =1,2$,
\\
$\epsilon^{01}=\epsilon^{\tau\sigma}=+1$,\\
$0\leq \sigma \leq \pi$.\\ The fields $X^a$ are dimensionless and periodic in
the target space with period
$2\pi$; $g_{ab}$ and $b_{ab}$ have dimension $[L]^2$. Their
conjugate momenta are
\begin{equation}
\Pi_a(\sigma,\tau)={1\over 2 \pi
\alpha^\prime}(g_{ab}\partial_\tau X^b -  b_{ab}
\partial_\sigma X^b)\ .
\label{moment}
\end{equation}
The configuration space  torus is defined
by a periodic lattice
\begin{equation}
\label{target} {\bf x}     \equiv  {\bf x} + 2\pi {\bf L}\ ,
\end{equation}
where
$\bf L $ can be decomposed in the basis $\{ {\bf e_a} \}$
\begin{equation}
\label{ll} {\bf L}= n^a {\bf e_a}\quad, \quad n^a \in Z\ .
\end{equation} The metric of the lattice is given by
\begin{equation}
\label{metr} g_{ab}= {\bf e_a.e_b}\ .
\end{equation}
The  metric and antisymmetric tensors which give  rise to
an EN lattice in the closed string sector with gauge group
${\cal G}
\times {\cal G}$ are characterised  by \cite{egrs}
\begin{equation} {\bf e_a} =  {1\over2}\sqrt{2\alpha^\prime}{\bf r_a}\ ,
\label{eroot}
\end{equation} where the vectors ${\bf r_a}$ form a basis of the root lattice
$\Lambda_{root}$ of $\cal G$ and are chosen  to be the simple roots with ${\bf
r_a.r_a}=2$.  The constant antisymmetric background field is
\cite{egrs}
\begin{equation}
 b_{ab} = +{\bf e_a.e_b}\ \hbox{for}\  a> b\ , \ = -{\bf e_a.e_b}\
\hbox{for}\   a< b\  ;
\quad b_{ab} =0\ \hbox{for}\   a=b\ .
\label{oldb}
\end{equation}
 It follows from these equations
that $(2/\alpha^\prime) g_{ab}   $ is the Cartan matrix of  $\cal G$  and
therefore,
\begin{equation}
\label{gquant}
{2 \over \alpha^\prime} g_{ab} \in Z\ .
\end{equation}
We also have
\begin{equation}
\label{bquant}
{2 \over \alpha^\prime} b_{ab} \in Z\ .
\end{equation}
The dual
basis ${\bf e^a}$ of
${\bf e_a}$ is
\begin{equation} {\bf e^a} =  2(2\alpha^\prime)^{-1/2}{\bf w^a}\ ,
\label{eweight}
\end{equation} where the ${\bf w^a}$ are the fundamental weights.
The left and right
momenta are given by:
\begin{eqnarray} {\bf p_R} &=& [({1\over2} m_a  - b_{ab} n^b
(2\alpha^\prime)^{-1}) {\bf e^a} + n^a (2\alpha^\prime)^{-1}{\bf e_a}]\ ,\ \
\nonumber \\ {\bf p_L} &=& [({1\over2} m_a  - b_{ab}
n^b(2\alpha^\prime)^{-1}) {\bf e^a} - n^a (2\alpha^\prime)^{-1} {\bf e_a}]\ .
\label{close}
\end{eqnarray}
One verifies that\\  {\it - a)} the left and right momenta
$\sqrt{2\alpha^\prime}{\bf p_R}$ and
$ \sqrt{2\alpha^\prime}{\bf p_L}$ lie on the weight lattice
$\Lambda_{weight}$ of $\cal G$\\ {\it - b)} the momenta
$\sqrt{2\alpha^\prime}{\bf p_R}$ and
$\sqrt{2\alpha^\prime}{\bf p_L}$ belong to the same conjugacy class:
$\sqrt{2\alpha^\prime}({\bf p_R}-{\bf p_L}) \ \in\
\Lambda_{root}$.  Thus the above choice for $g_{ab} $ and $ b_{ab}$ does
indeed characterise a toroidal compactification on an EN lattice
\cite{en}.
The boundary condition for open strings in the compact dimensions is,
\begin{equation} (g_{ab}\partial_\sigma X^b -  b_{ab}
\partial_\tau X^b)\delta X^a
\vert^{\sigma=\pi}_{\sigma=0} =0 \ ,
\end{equation}
which can be satisfied by either Dirichlet conditions
\begin{equation}
\label{dirichlet}
\delta X^a
\vert^{\sigma=\pi}_{\sigma=0} =0 \ ,
\end{equation}
or generalised Neumann conditions
\begin{equation}
\label{neumann}
[g_{ab}\partial_\sigma X^b -  b_{ab}
\partial_\tau X^b]
\vert^{\sigma=\pi}_{\sigma=0} =0 \ .
\end{equation}
Both conditions yield the same
open string spectrum, a consequence of the E-duality generalisation of
T-duality at the enhanced symmetry points of toroidal compactification.
In terms of the Dirichlet winding ${\bf L}$ in Eq.(\ref{ll}) and of the Neumann
momentum
${\bf p}$, we have
\begin{eqnarray}
\label{massD}
\alpha^\prime m^2&= &{1\over\alpha^\prime}{\bf L.L}
+  N -1\ ,\\
\label{massN}
\hbox{or}\qquad \alpha^\prime m^2&=& \alpha^\prime{\bf p.p}
+  N -1\ , \\
\label{duality}
\hbox{with}\quad\qquad  {\bf L}&=&\alpha^\prime {\bf p} \ .
\end{eqnarray}
The last equation Eq.(\ref{duality}) expresses the E-duality for open
strings and shows that $\sqrt {2\alpha^\prime} {\bf p}$ belongs to the root
lattice.

\section{ Group-invariant twist operator}

For Neumann boundary conditions Eq.(\ref {neumann})
we have
\begin{eqnarray} X^a(\sigma,\tau)=x^a  &+& g^{ac} b_{cb} B^b\sigma +
B^a \tau +\sum_{n\neq 0} D_n^a\cos n\sigma
\exp(-in\tau)\nonumber\\ &-&\sum_{n\neq 0} ig^{ac} b_{cb}D_n^b\sin n\sigma
\exp(-in\tau)\ .
\label{positionN}
\end{eqnarray} The conjugate momentum Eq.(\ref{moment}) then  takes
the form
\begin{eqnarray} \Pi_a(\sigma,\tau)&=&{1\over 2 \pi
\alpha^\prime}[(g_{ab}
 -  b_{ac}g^{cd}b_{db})B^b \nonumber\\&-&i\sum_{n\neq 0}n(g_{ab}
 -  b_{ac}g^{cd}b_{db}) D_n^b\cos n\sigma\exp(-in\tau)]\ .
\label{momentN}
\end{eqnarray} To compute the commutation relations between the
operators
$ x^a$, $B^a$ and $D_n^a$ it is convenient to introduce  the tensor
$e_{ab} \equiv g_{ab}+b_{ab}$ and to define the E-dual metric
$G^{ab}$ and antisymmetric tensor
$ B^{ab}$ by $(e^{-1})^{ab}\equiv G^{ab}+B^{ab}$
\begin{equation} (g_{ac}+b_{ac})(G^{cb} + B^{cb})=\delta^b_a\ .
\end{equation}  This is equivalent to
\begin{eqnarray} g_{ab} -  b_{ac}g^{cd}b_{db} = (G^{-1})_{ab} \ ,&&\nonumber \\
g^{ac}b_{cb} = -  B^{ac }(G^{-1})_{cb}\ .&&
\label{inverse}
\end{eqnarray} From Eqs.(\ref{positionN}), (\ref{momentN}) and
(\ref{inverse}) we get
 \begin{equation} B^a = (2\alpha^\prime)  G^{ab} p_b\quad ,\quad
D_n^a= i\sqrt{2\alpha^\prime} {\alpha^a_n\over n}\ ,
\end{equation} where the eigenvalues of the  momentum
operators            $ p_b$ are
integers
$m_b$. We rewrite
$X^a$ in terms of the E-dual variables as
\begin{eqnarray} X^a = x^a &-& 2\alpha^\prime  B^{ab } p_b \sigma
+ 2\alpha^\prime G^{ab } p_b
\tau +i\sqrt{2\alpha^\prime }\sum_{n\neq 0} {1\over n}
\alpha^a_n\cos n\sigma
\exp(-in\tau)\nonumber\\ &-&\sqrt{2\alpha^\prime}\sum_{n\neq 0} {1 \over n}
B^{ac }(G^{-1})_{cb} \alpha^b_n\sin n\sigma
\exp(-in\tau)\label{positionnew}\ ,
\end{eqnarray}
where
\begin{equation}
\qquad [\alpha^a_m,\alpha^b_n] = G^{ab} m \delta_{m+n,0}\ .
\end{equation}

The conventional twist operator is\begin{equation}
\label{twist}
\Omega = e^{i\pi N}\quad ,\quad  N=\sum_{n>0}
(G^{-1})_{ab}\alpha^a_{-n}\alpha^b_n \ ,
\end{equation}
and gives
\begin{equation}
\label{wrong}
\Omega X^a(\sigma; B)\Omega^{-1} = X^a(\pi- \sigma; -B)
+\pi(2\alpha^\prime) B^{ab} p_b\ .
\end{equation}
The matix elements  $(2\alpha^\prime)B^{ab}$ are, for EN lattices, $\pm1$or
$0$ and the eigenvalues of the last term in Eq.(\ref{wrong})  shifts $ X^a(\pi-
\sigma; -B) $ to non equivalent lattice points. This effect is corrected by
introducing a twist operator which is  group invariant:
\begin{equation}
\Omega^\prime = \Omega e^{i\pi \alpha^\prime{\bf p}^2}= e^{i\pi (N
+\alpha^\prime {\bf p}^2)}\ ,
\end{equation} where {\bf p} is the Neumann momentum Eq.(\ref{duality}) and
${\bf p}^2= p_a G^{ab}p_b$.

Eq.(\ref{wrong}) becomes
\begin{equation}
\label{wright}
\Omega^\prime X^a(\sigma; B)\Omega^{\prime -1} = X^a(\pi- \sigma; -B)
+\pi (2\alpha^\prime)(e^{-1})^{ab} p_b\ .
\end{equation}

The matrix elements  $(e^{-1})^{ab}$ are even integers and hence
 induce a shift to equivalent lattice points. Thus the group
invariant operator
$(1+\Omega^\prime)/2$ is a valid projection operator onto  unoriented open
strings. Note that for $SU(2)$ $B^{ab}$ is zero and Eq.(\ref{wrong}) makes
sense. Nevertheless, the group invariant operator still induces a shift to
equivalent lattice points.

\section{Torus volumes}
Using Eqs.(\ref{metr}) and Eq.(\ref{eroot}),  we can compute $\sqrt{g}$ for EN
lattices. We obtain
\begin{equation} \sqrt{g} =  \vert {\bf  e_a}{}_1\wedge{\bf
 e_a}{}_2\wedge....\wedge{\bf  e_a}{}_{\mbox {\scriptsize d}}\vert =
   \left({{\alpha^{\prime} \over 2}}\right)     ^{d/ 2}  V_{root}\ ,
\end{equation} where $V_{root}$ is the volume of the elementary cell of the
${\cal G}$ root lattice. It is equal to $ \sqrt {\cal N}$ (see footnote 8).
We thus
get
\begin{equation}
\label{vg}
 (2 \pi)^d \sqrt{g} =  (2\pi\alpha^\prime{}^{1/2})^d { \sqrt {\cal N}\over
2^{d/2}}\ .
\end{equation}
To evaluate $\sqrt  e$, we notice using Eqs.(\ref{metr}),  (\ref{oldb}) that
${e_{ab}}$ is a triangular matrix. The determinant is given by
the product of the diagonal elements
$e_{aa}= (\alpha^{\prime} /2) {\bf r_a. r_a}= \alpha^{\prime}$. We thus find
\begin{equation}
\label{eg} (2 \pi)^d \sqrt{e} = (2\pi\alpha^\prime{}^{1/2})^d\ =V_{sd}\ ,
\end{equation} which corresponds to the self-dual volume.

Therefore
\begin{equation}
\label{fuse1}  \sqrt{e \over g} ={2^{d/ 2} \over \sqrt
{\cal N}}\ .
\end{equation}

\end{document}